\documentclass{article}
\usepackage[utf8]{inputenc}
\usepackage[margin=1in]{geometry}
\usepackage{rotating}
\usepackage[numbered]{bookmark}
\usepackage[FIGTOPCAP]{subfigure}
\usepackage{placeins}

\usepackage{amsfonts, amsmath, amsthm, amssymb, bm}
\usepackage{aligned-overset}

\usepackage[title]{appendix}

\usepackage[ruled,vlined]{algorithm2e}

\usepackage[english]{babel}
\newtheorem{theorem}{Theorem}
\newtheorem{proposition}{Proposition}
\newtheorem{lemma}{Lemma}
\newtheorem{definition}{Definition}

\newcommand{\indep}{\;\rotatebox[origin=c]{90}{$\models$}\;}

\newcommand{\E}[1]{\mathbb{E}\left[#1\right]}

\renewcommand{\P}[1]{\mathbb{P}\left[#1\right]}

\makeatletter
\newcommand{\oset}[2]{%
  {\mathop{#2}\limits^{\vbox to 1.75\ex@{\kern-\tw@\ex@
   \hbox{\scriptsize #1}\vss}}}}
\makeatother

\usepackage{bbm}
\newcommand{\I}[1]{\mathbbm{1}\left[#1\right]}

\usepackage[table]{xcolor}

\usepackage{float}
\usepackage{booktabs, multirow, ragged2e, array}

\newcolumntype{C}[1]{>{\centering\let\newline\\\arraybackslash\hspace{0pt}}m{#1}}

\usepackage[
    backend=biber,
    style=numeric-comp,
    sortcites,
    sorting=none,
    giveninits=true,
    natbib,
    maxbibnames=99,
    doi=false,isbn=false,url=false,eprint=false
]{biblatex}
\hypersetup{hidelinks}

\title{Searching for subgroup-specific associations while controlling the false discovery rate}

\author{Matteo Sesia\thanks{Department of Data Sciences and Operations, University of Southern California, Los Angeles, CA 90089, USA.}\, and Tianshu Sun\footnotemark[1]}
\date{\today}

\begin{document}

\maketitle

\begin{abstract}
This paper introduces an innovative method for conducting conditional independence testing in high-dimensional data, facilitating the automated discovery of significant associations within distinct subgroups of a population, all while controlling the false discovery rate. This is achieved by expanding upon the model-X knockoff filter to provide more informative inferences. Our enhanced inferences can help explain sample heterogeneity and uncover interactions, making better use of the capabilities offered by modern machine learning models. Specifically, our method is able to leverage any model for the identification of data-driven hypotheses pertaining to interesting population subgroups. Then, it rigorously test these hypotheses without succumbing to selection bias. Importantly, our approach is efficient and does not require sample splitting. We demonstrate the effectiveness of our method through simulations and numerical experiments, using data derived from a randomized experiment featuring multiple treatment variables.
\end{abstract}


\section{Introduction} \label{sec:intro}

\subsection{Background and motivation}

Contemporary data sets are growing in both size and diversity, often incorporating observations from various environments or diverse populations characterized by a multitude of distinct patterns and behaviours.
This complexity offers many opportunities to statisticians. For instance, it can help alleviate issues of collinearity \citep{peterson2019genome}, enhance the robustness of predictive models to distribution shifts \citep{berg2019reduced}, and aid in the identification of causal associations by filtering out spurious correlations \citep{peters2016causal,buhlmann2020invariance}. 
However, such heterogeneous data also pose new challenges, highlighting some limitations of existing analysis methods that rely on rigid population-wide hypothesis testing.

This paper focuses on the problem of identifying significant conditional associations between an outcome of interest and many possible explanatory variables.
For this purpose, we adopt and expand the {\em model-X} framework proposed by \citet{candes2018}, which does not require assuming a parametric regression model for the outcome.
Within this context, we contend that population-wide hypothesis testing is not fully satisfactory when dealing with heterogeneous samples. Such testing may yield technically valid discoveries, but it leads to findings that are not necessarily the most practically valuable, particularly when the objective is to guide data-driven decisions involving diverse individuals who exhibit varying behaviours.
This concern serves as the driving force behind our development of a novel methodology, one that can provide more informative model-X inferences tailored to interesting subgroups of individuals.

Our work is partly motivated by the analytical challenges faced by e-commerce and social media enterprises, which have access to extensive data \citep{george2014big} and are frequently able to monitor customer responses to randomized interventions \citep{aral2012identifying}.
These experiments often entail multiple treatment variables, and their joint distribution can be intricate, and possibly contingent on many covariates \citep{aral2012identifying}. 
For instance, consider a scenario within social networks where the ``things in common'' are displayed to encourage friendship formation \citep{sun2020displaying}. In this case, users who hover on another user's profile are randomly presented with some shared features (e.g., having attended the same school, expressing similar interests, etc.), with the objective of discerning which interventions facilitate the creation of new connections.
Alternatively, researchers in an e-commerce company may wish to explore the efficacy of different targeted marketing strategies, leveraging insights from the previous purchases and browsing histories of current customers to make personalized product recommendations.

In both of the aforementioned scenarios, there may exist numerous significant associations between the outcome of interest and various explanatory variables. Given a sufficiently large sample size, any sensible test should be able to identify these associations. Nevertheless, converting these findings into actionable business insights can prove to be a difficult task since these associations may hold little relevance for the majority of individuals in the population.

Consider, for instance, a subscription box service offering an array of products to its subscribers on a monthly basis. 
The company's objective is to understand how various promotional strategies impact customer retention and product preferences. Their data collection encompasses a range of promotional tactics, such as discounts, exclusive items, and early access.
Upon analysis, an association may emerge between providing exclusive outdoor adventure gear (e.g., backpacking food) as part of the subscription box and an increase in customer retention. 
Nonetheless, this association probably holds relevance only for a {\em minority} of users who belong to a subgroup characterized by specific demographic attributes (e.g., younger adults), geographical features (e.g., proximity to mountains) or who have purchase histories of related equipment (e.g., hiking gear).
Therefore, the analysts would likely find it helpful to have at their disposal methods that can not only detect significant associations, but also automatically pinpoint the population subgroups for which the discoveries are {\em relevant}.

Although the potential impacts of this statistical problem are far-reaching, and wealth of related literature has already been written, we contend that there remains an unaddressed void in the current methodology that necessitates the development of a novel approach.
On one hand, there are many useful data mining algorithms and machine learning models that can discover subgroups of individuals exhibiting coherent behaviours \citep{wrobel1997algorithm,novak2009supervised}, detect interactions \citep{bien2013lasso,tian2014simple}, or estimate heterogeneous effects in regression and causal inference \citep{verbeke1996linear,caron2022estimating}. 
However, such methods are usually not designed to offer precise statistical guarantees in finite samples.
On the other hand, there is the {\em model-X} framework of \citet{candes2018}, which makes it possible to rigorously test conditional independence hypotheses under finite-sample type-I error control, while taking advantage of any model to powerfully discover significant associations.
However, existing model-X methods are currently limited in the analysis of heterogeneous data because they can only test rigid hypotheses defined over the entire population.
Such discoveries may be useful to screen out completely irrelevant predictors but are not directly informative about the particular effects of any variables across diverse subgroups of individuals, and they cannot be used to test for interactions.


\subsection{Preview of our contribution} \label{sec:preview}

This paper presents an extension of the model-X knockoff filter \citep{candes2018} that enhances the informativeness of inferences derived from heterogeneous data and in the presence of interactions.
Our novel method accomplishes two key objectives: (1) it harnesses the power of any machine learning model to automatically identify interesting conditional independence hypotheses specific to well-defined population subgroups, and (2) it rigorously tests these data-driven hypotheses while controlling the false discovery rate \citep{benjamini1995} in finite samples.
Notably, our method efficiently utilizes the same data for both tasks, avoiding selection bias \citep{assmann2000subgroup,cook2004subgroup,wager2018estimation} without wasteful sample splitting.
We call this method the {\em subgroup-selective knockoff filter}.

\begin{figure}[!htb]
  \centering
  \includegraphics[width=\textwidth]{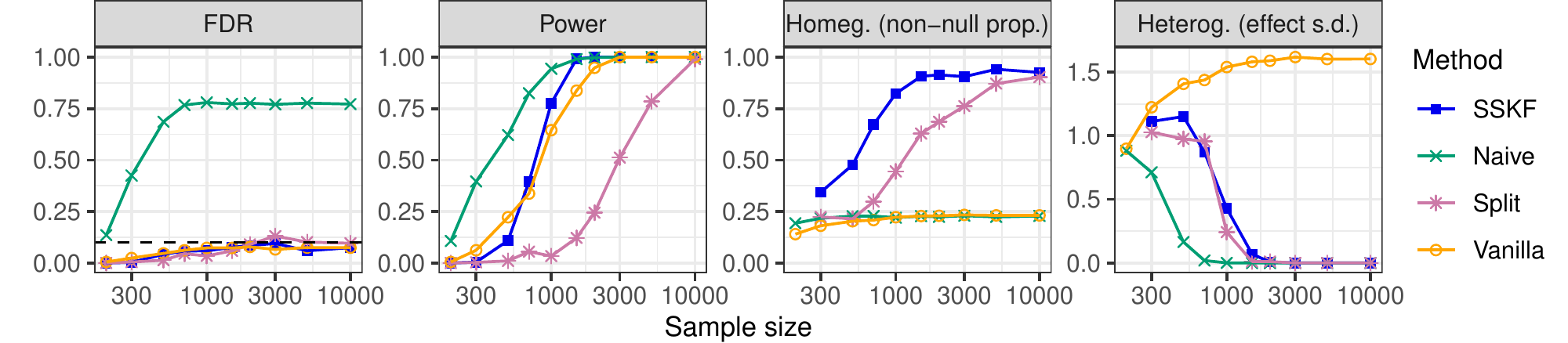}\\[-0.5em]
  \caption{Performance of the subgroup-selective knockoff filter (SSKF) and benchmark methods on synthetic data. The informativeness of the discoveries is quantified by the homogeneity of the corresponding subgroups (higher is better) and the heterogeneity of the underlying true effects in the data-generating model (lower is better). The nominal false discovery rate (FDR) level is 0.1. }
  \label{fig:experiment-heterogeneous-1}
\end{figure}

Figure~\ref{fig:experiment-heterogeneous-1} gives a preview of our method's performance. These numerical experiments are conducted using simulated data that consist of multiple explanatory (or ``treatment'') variables and an outcome generated from a parametric model, which we refer to as the true ``causal'' model.
This underlying model is intentionally structured so that the outcome is influenced by a distinct subset of variables for different subgroups of individuals, as defined by specific observable covariates. To be more precise, the data generation model is a linear model that incorporates interaction terms between the explanatory variables and the covariates. Consequently, each variable has a unique effect on each individual, with this effect often equating to zero and generally contingent upon certain covariates.
The analyst's objective is to discover, with the highest possible precision, which variables exert a non-zero effect within particular subgroups.

As the ground-truth model is known, it is possible to verify which discovered associations are not spurious.
Further, it is also feasible to gauge the informativeness of any findings.
This can be achieved by comparing the sub-populations identified by the rejected hypotheses and the subgroups of individuals characterized by non-zero treatment effects within the data-generating model.
To be more specific, we introduce two distinct metrics for assessing informativeness: one metric quantifies the homogeneity of the true effects within the reported subgroups, while the other metric quantifies their heterogeneity in a complementary manner.
We refer to Section~\ref{sec:experiments} for further information about the setup of these numerical experiments.

The first alternative (benchmark) approach considered in Figure~\ref{fig:experiment-heterogeneous-1} is the vanilla model-X knockoff filter \citep{candes2018}.
 This method can be powerful in discovering population-wide conditional associations and it reliably controls the false discovery rate.
 However, it is unable to pinpoint the subgroups in which the significant variables indeed possess non-zero effects.
The limitations of this approach are evident in Figure~\ref{fig:experiment-heterogeneous-1}, where the informativeness metrics remain relatively unchanged as the sample size increases.

In contrast, the subgroup-selective knockoff filter progressively uncovers more informative discoveries as the data set grows in size, ultimately enabling precise identification of subgroups with non-zero effects. 
Additionally, Figure~\ref{fig:experiment-heterogeneous-1} illustrates the strength of our method by comparing it to an intuitive ``data splitting'' benchmark. 
This benchmark involves applying the two distinct modules of the subgroup-selective knockoff filter (i.e., learning and testing) using separate subsets of the data obtained through random sample splitting. While this simpler approach also provides valid and informative inferences, it is notably less powerful than our method.

Finally, Figure~\ref{fig:experiment-heterogeneous-1} evaluates the performance of the subgroup-selective knockoff filter against a naive benchmark that differs from our method in that it does not carefully protect against possible selection bias. Unsurprisingly, the naive benchmark results in an excess of spurious discoveries.

\subsection{Related work} \label{sec:related}

Our work builds upon the model-X version \citep{candes2018} of the knockoff filter \citep{barber2015controlling}.
Prior research studied the problem of constructing (approximately) valid {\em knockoff} variables for different types of data \citep{sesia2019,gimenez2019knockoffs,jordon2018knockoffgan,romano2019,bates2020metropolized}; therefore, in this paper we  will assume that  suitable knockoffs are available for the application of interest.
Others have studied the robustness of the knockoff filter to approximations in construction of the knockoffs \citep{fan2019rank,barber2020robust} and its power \citep{liu2019power,wang2021high,spector2022powerful,katsevich2022power}, providing additional support for this framework.

This is the first paper in the model-X literature to seek subgroup-specific inferences for heterogeneous data.
Our work is most closely related to \citet{li2021searching}, which proposed an extension of the knockoff filter to analyze data collected from different environments, and argued that robust conditional associations in that context can be seen as relatively good proxies for causal inferences.
While we take some inspiration from \citet{li2021searching}, we study a different problem.
In fact, \cite{li2021searching} considered each individual as belonging to a distinct and {\it known} population, and their methods cannot test hypotheses corresponding to adaptively discovered subgroups.

The subgroup-selective knockoff filter is flexible and can learn which subgroups are likely to be affected by each variable using any model. 
This connects our work to the rich literature on subgroup analysis or the estimation of heterogeneous effects in regression and causal inference \citep{verbeke1996linear,zhao2012estimating,athey2016recursive,yao2018representation,moore2019linear,kunzel2019metalearners,hahn2020bayesian,nie2021quasi,cui2023estimating}; see also \citet{caron2022estimating} for a recent review.
The learning module of our method can take advantage of any such models, which may be either parametric or non-parametric, before the subsequent testing module translates their output into precise statistical inferences.


For simplicity, this paper focuses on learning using generalized linear models with interaction terms, as opposed to more complex approaches which may for example involve random forests \citep{wager2018estimation}, Bayesian additive regression trees \citep{hill2011bayesian}, or neural networks \citep{louizos2017causal}.
Interaction terms are relatively simple to explain in the context of (generalized) linear models, and they are an important source of sample heterogeneity because they can result in some variables having much larger effects in subgroups with certain covariates.
There are numerous works relating interactions to sample heterogeneity in regression or causal inference \citep{verbeke1996linear,moore2019linear,hahn2020bayesian,nie2021quasi}, and several techniques have been developed to discover interactions using high-dimensional data \citep{bien2013lasso,tian2014simple,wen2014bayesian,lim2015learning,yu2019reluctant,tibshirani2020pliable}.
We will demonstrate the use of the subgroup-selective knockoff filter in combination with some of these techniques. However, on their own, the latter would rely on different assumptions and provide different guarantees compared to our method.

\subsection{Outline of this paper}

Section~\ref{sec:preliminaries} introduces useful notation and recalls the relevant technical details of the model-X knockoff filter from \citet{candes2018}.
Section~\ref{sec:methods} states our problem and presents our method.
Section~\ref{sec:empirical} demonstrates the use of our method empirically.
Section~\ref{sec:discussion} concludes with some ideas for future work.
The Appendices contains additional methodological details, numerical results, and mathematical proofs.
The Appendices also extends our novel method to test {\em partial conjunctions} \citep{friston2005conjunction,benjamini2008screening} of our subgroup-specific hypotheses, complementing the types of inferences considered in this paper.

\section{Technical preliminaries} \label{sec:preliminaries}

\subsection{Basic notation and setup} \label{sec:notation}

We consider a data set comprising $n$ individual observations of triplets $(X^{i}, Y^{i}, Z^{i})$. For each individual $i \in [n] = \{1,\ldots,n\}$, $Y^{i} \in \mathbb{R}$ is the outcome of interest (which may be either numerical or categorical), ${X^{i} = (X^{i}_1, \ldots, X^{i}_p)  \in \mathbb{R}^p}$ describes $p$ explanatory variables, and $\smash{Z^{i} = (Z^{i}_1, \ldots, Z^{i}_m) \in \mathbb{R}^m}$ represents $m$ additional covariates.
Let us denote as  $\mathbf{X} \in \mathbb{R}^{n \times p}$, $\mathbf{Y} \in \mathbb{R}^{n \times 1}$, and $\mathbf{Z} \in \mathbb{R}^{n \times m}$ the data matrices collecting the observations for all individuals. 
Without loss of generality, the data distribution can be factored as $P_{X,Y,Z}(\mathbf{X}, \mathbf{Y}, \mathbf{Z}) = P_{Z}(\mathbf{Z}) \cdot P_{X \mid Z} (\mathbf{X} \mid \mathbf{Z}) \cdot P_{Y \mid X, Z}(\mathbf{Y} \mid \mathbf{X}, \mathbf{Z})$.
In the following, $P_{X \mid Z}$ will be assumed to be known in order to generate knockoffs, while $P_{Z}$ and $P_{Y \mid X, Z}$ may remain completely arbitrary and unknown.

The assumption that $P_{X \mid Z}$ is known may not be justified in every application, but it is often a reasonable approximation and is well-suited for the analysis of data from randomized experiments \citep{nair2023randomization} or genome-wide association studies \citep{sesia2019}. Further, it is feasible to verify the appropriateness of this assumption in practice \citep{romano2019}.

Intuitively, the goal is to shed some light onto how $Y$ depends on $X$ given $Z$; this aim will be stated more precisely later in terms of testing suitable conditional independence hypotheses.
For simplicity, we will assume that different individuals are independent of one another.

\subsection{Relevant background on conditional testing with knockoffs} \label{sec:knockoffs-review}

The model-X problem studied by \citet{candes2018} can be thought of as testing, for all $j \in [p]$, whether $X_j$ is associated with $Y$ given $Z$ and all variables excluding $X_j$ (i.e., $X_{-j}$); that is, whether the following null hypothesis holds true:
\begin{align} \label{eq:null-hyp}
  \mathcal{H}_{0,j} : Y \indep X_j \mid X_{-j}, Z.
\end{align}
We refer to Appendix~\ref{app:standard-hypotheses} for further details regarding the interpretation of~\eqref{eq:null-hyp} and its connections to parametric inference for generalized linear models and causal inference.

The model-X knockoff filter \citep{candes2018} is designed to test $\mathcal{H}_{0,j}$ in~\eqref{eq:null-hyp} for all $j \in [p]$ while controlling the false discovery rate---the expected proportion of falsely rejected hypotheses.
This is an especially useful error rate for analyses in which many discoveries are expected.

The key ingredient for the knockoff filter are the knockoffs, which we denote as $\tilde{X}$.
These are synthetic variables constructed by the statistician as a function of $X$ and $Z$ without looking at $Y$, so that $\tilde{X} \indep Y \mid X,Z$.
Further, the knockoffs are designed to be exchangeable with $X$ in the joint distribution of $(X,\tilde{X})$ conditional on $Z$.
That is, if $[ \mathbf{X}, \tilde{\mathbf{X}} ] \in \mathbb{R}^{n \times 2p}$ is the matrix obtained by concatenating $\mathbf{X}$ with the corresponding $\tilde{\mathbf{X}} \in \mathbb{R}^{n \times p}$ and, for any $j \in [p]$, the matrix $[ \mathbf{X}, \smash{\tilde{\mathbf{X}}]_{\mathrm{swap}(j)}}$ is obtained by swapping the $j$-th columns of $\mathbf{X}$ and ${\tilde{\mathbf{X}}}$, then
\begin{align} \label{eq:knock_cond_1}
  \big[ \mathbf{X}, \tilde{\mathbf{X}} \big]_{\mathrm{swap}(j)} \mid \mathbf{Z}
  \; \overset{d}{=} \; \big[ \mathbf{X}, \tilde{\mathbf{X}} \big] \mid \mathbf{Z}, \qquad \forall j \in [p],
\end{align}
where the symbol $\overset{d}{=}$ denotes equality in distribution.
Thus, the only meaningful difference between $X_j$ and $\tilde{X}_j$ may be the lack of conditional association of the latter with~$Y$.

Constructing knockoffs that satisfy~\eqref{eq:knock_cond_1} requires knowledge of $P_{X \mid Z}$, as anticipated in Section~\ref{sec:notation}, and it can be computationally involved. However, the problem is well-studied; see Section~\ref{sec:related}. Therefore, this paper assumes that suitable knockoffs are available for the data at hand.

The purpose of the knockoffs is to serve as negative control variables within a predictive model for $Y$ given $X,\tilde{X}$ and $Z$.
This model is utilized compute importance measures $T_j$ and $\tilde{T}_j$ for each $X_j$ and $\tilde{X}_j$.
Any model can be employed for this purpose, as long as swapping $X_j$ with $\tilde{X}_j$ only results in $T_j$ being swapped with $\tilde{T}_j$.
A typical choice is to fit a (generalized) lasso model \citep{tibshirani1996regression} and define the importance measures as the absolute values of the scaled regression coefficients for $X$ and $\tilde{X}$, after tuning the regularization via cross-validation.
Then, $T_j$ and $\tilde{T}_j$ are combined pairwise for each $j$ into anti-symmetric statistics $W_j$; i.e., $W_j = T_j - \smash{\tilde{T}_j}$.
This ensures that the signs of the $W_j$ are mutually independent coin flips for all $j$ corresponding to a true $\mathcal{H}_{0,j}$, while larger and positive values provide evidence against the null.
Letting $\epsilon \in \{-1,+1\}^p$ be an independent random vector such that $\epsilon_j = +1$ if $\mathcal{H}_{0,j}$ is false and $\P{\epsilon_j = +1 } = 1/2$ otherwise, then it can be proved that $W$ satisfies the {\em flip-sign} property:
\begin{align} \label{eq:flip-sign}
  W \mid \mathbf{Z} \overset{d}{=} W \odot \epsilon \mid \mathbf{Z},
\end{align}
where $\odot$ indicates element-wise multiplication.
Intuitively, the sign of $W_j$ gives rise to a conservative binary p-value $p_j$ for $\mathcal{H}_{0,j}$.
That is, $p_j = 1/2$ if $W_j>0$ and $p_j = 1$ otherwise, as long as $\mathcal{H}_{0,j}$ is true.
Finally, a rejection rule for~\eqref{eq:null-hyp} can be obtained by applying a sequential testing procedure to these p-values, in the order defined by the absolute values of $W$.
As each of these p-values contains a single bit of information, an appropriate sequential testing procedure is the knockoff filter (or selective SeqStep+ test) of~\cite{barber2015controlling}, which can compute an adaptive threshold for $W_j$ controlling the false discovery rate below any desired threshold.

\section{Methodology} \label{sec:methods}

This section is organized as follows.
Section~\ref{sec:individualized-hyp} defines our subgroup-specific conditional hypotheses, extending the population-wide hypotheses defined in~\eqref{eq:null-hyp}.
Section~\ref{sec:invariant-subgroups} gives general guidelines for learning adaptive subgroups that can lead to informative hypotheses without introducing selection bias in our subsequent tests.
Section~\ref{sec:learn-interactions} outlines a concrete approach for learning data-driven subgroups using parametric regression models with interaction terms.
Section~\ref{sec:skf} explains the core of our method, which can test adaptively discovered subgroup-specific conditional hypotheses while controlling the false discovery rate.
Section~\ref{sec:powerful-statistics} provides a detailed example of how to implement our method using powerful test statistics.

\subsection{Subgroup-specific conditional hypotheses} \label{sec:individualized-hyp}

Standard conditional testing is not fully satisfactory when analyzing heterogeneous data.
In fact, rejecting  $\mathcal{H}_{0,j}$ in \eqref{eq:null-hyp} informs us that $X_j$ may have {\em some} conditional association with $Y$ in the population, but it sheds no light on the possible heterogeneity of this relation.

Consider for example a simple mixture population with $p=2$ and $m=1$, in which $Z_1=1$ for half of the individuals and $Z_1=0$ for the others.
Suppose $X_1$ and $X_2$ are independent standard normal and $Y \sim \mathcal{N}(Z_1X_2\beta_2, 1)$, for some coefficient $\beta_2 \in \mathbb{R}$. Then, \eqref{eq:null-hyp} is true if and only if $\beta_2=0$. Thus, testing $\mathcal{H}_{0,j}$ for $j \in \{1,2\}$ may be helpful to distinguish $X_2$ from $X_1$, but it does not inform us about the interaction with $Z_1$.
This motivates the more flexible hypothesis testing framework presented below, which will allow us to automatically discover whether a significant association holds only within a specific population subgroup.
Continuing with the previous example, our method will be able to discover that $X_2$ is important within the population subgroup with $Z_1=1$, and it will report no evidence of such association among individuals with $Z_1=0$. Crucially, there will be no need to know in advance that $X_2$ interacts with $Z_1$.

For any variable index $j \in [p]$ and constant $G \in \mathbb{N}$, consider a {\em fixed} function $\psi_j : \mathbb{R}^m \mapsto [G]$ that partitions the covariate space $\mathbb{R}^m$ into $G$ disjoint regions, each identifying a distinct population {\it subgroup}.
For example, in the case of the toy model described above, the most informative partition function would intuitively be $\psi_1 = 1$ and $\psi_2 = 1+\I{Z_1=1}$, where $\I{\cdot}$ is the indicator function.
Note that the number of possible subgroups can generally vary implicitly across different variables, and $G$ simply refers to the largest partition size across all $j \in [p]$.
In fact, the function $\psi_1$ in the previous example defines a single trivial subgroup encompassing the whole population, which is a reasonable choice in this case since $X_1$ is unimportant for all individuals.

To extend the standard population-wide conditional testing framework~\eqref{eq:null-hyp}, we propose to study {\em subgroup-specific} conditional hypotheses of the form 
\begin{align} \label{eq:null-hyp-env}
  \mathcal{H}^{(g)}_{0,j} : Y^{(g)} \indep X^{(g)}_j \mid X^{(g)}_{-j}, Z^{(g)},
\end{align}
for each $j \in [p]$ and $g \in [G]$, where $\smash{(X^{(g)},Y^{(g)},Z^{(g)})}$ denotes a random sample from the distribution $P_{X,Y,Z}$ restricted to $\smash{\{z \in \mathbb{R}^m : \psi_j(z) = g\}}$, which we call $\smash{P^{(g)}_{X,Y,Z}}$.
Note that the dependence of the group $g$ on the variable index $j$ is not shown explicitly to shorten the notation.
Considering~\eqref{eq:null-hyp-env} for all $j \in [p]$ and $g \in [G]$ gives a multiple testing problem with $m G$ hypotheses, if the $\psi_j$ are fixed.
In the special case of $G=1$, \eqref{eq:null-hyp-env} reduces to~\eqref{eq:null-hyp}.
In general, however, a rejection of~\eqref{eq:null-hyp-env} is more informative than one of~\eqref{eq:null-hyp}, because \eqref{eq:null-hyp-env} is implied by \eqref{eq:null-hyp} but the converse is not true.
In particular, rejecting~\eqref{eq:null-hyp-env} can help us to pinpoint the effect of $X_j$ to subgroup $g$.

An appropriate choice of $\psi_j$ is therefore essential to ensure we test interesting hypotheses.
In fact, \eqref{eq:null-hyp-env} can be more informative than~\eqref{eq:null-hyp} only if the subgroups are relatively homogeneous in their associations with $Y$.
Thus, unless prior information about $P_{Y \mid X,Z}$ is available, $\psi_j$ should be data-driven.
As mentioned in Section~\ref{sec:intro}, there exist many subgroup discovery algorithms, but one must be careful that allowing the hypotheses to be random may create the risk of selection bias in the subsequent tests.
In particular, if the same data utilized to select the subgroups were re-used naively to test the hypotheses in \eqref{eq:null-hyp-env}, the type-I errors might be inflated \citep{wager2018estimation}, consistently with Figure~\ref{fig:experiment-heterogeneous-1}.
Sample splitting could of course circumvent this issue, but it is unnecessarily wasteful.

\subsection{Knockoff-invariant subgroups} \label{sec:invariant-subgroups}

We now begin to present a novel method that can powerfully test subgroup-specific conditional hypotheses~\eqref{eq:null-hyp-env} adaptively learnt from the data, while controlling the false discovery rate.
The key idea is that one can select informative subgroups and then test the corresponding hypotheses while avoiding an uncontrollable inflation of the type-I errors, as long as the learning component of the analysis satisfies a suitable invariance property.
This section defines the necessary invariance property and explains how to achieve it in practice.

Let $\tilde{\mathbf{X}} \in \mathbb{R}^{n \times p}$ be a matrix of knockoffs for $\mathbf{X}$, generated with existing techniques.
For convenience, we refer to the knockoff-augmented data as $\mathcal{D} = (\mathbf{X}, \tilde{\mathbf{X}}, \mathbf{Y}, \mathbf{Z})$.
For any $j \in [p]$, let $\smash{\hat{\psi}_j: {\mathbb{R}^m} \mapsto {\mathbb{N}}}$ be a random function whose parameters may depend on $\mathcal{D}$ as well as on an independent random matrix $\mathbf{V}$ generated by the analyst, which is specified precisely below; i.e., $\smash{ \hat{\psi}_j(z; [\mathbf{X}, \tilde{\mathbf{X}}], \mathbf{Y}, \mathbf{Z}, \mathbf{V}) \in \mathbb{N} }$.
In the following, we will focus on a special class of invariant partition functions that are allowed to depend on the data in $\mathcal{D}$ but cannot make use of any knowledge about which variables are real and which are knockoffs.

\begin{definition} \label{def:knockoff-invariant-function}
Let $\mathbf{V} \in \{0,1\}^{n \times p}$ be i.i.d.~Bernoulli random variables generated by the analyst.
Denote as $[\mathbf{X},\tilde{\mathbf{X}}]_{\mathrm{swap}(\mathbf{V})} \in \mathbb{R}^{n \times 2p}$ the concatenation of $\mathbf{X}$ and $\tilde{\mathbf{X}}$, with the $i$-th observation of $X_j$ swapped with its knockoff if and only if $V_{ij} = 1$.
Then, we say that a vector-valued partition function $\hat{\psi} : \mathbb{R}^m \mapsto [G]^p $ is knockoff-invariant if it can be written as
$$ \hat{\psi}_j(z; [\mathbf{X}, \tilde{\mathbf{X}}], \mathbf{Y}, \mathbf{Z}, \mathbf{V}) = \hat{\psi}^0_j(z; [\mathbf{X}, \tilde{\mathbf{X}}]_{\mathrm{swap}(\mathbf{V})}, \mathbf{Y}, \mathbf{Z}),$$
for all $j \in [p]$, $z \in \mathbb{R}^m$, and for some fixed $\hat{\psi}^0: \mathbb{R}^m \to [G]^p$ whose parametrization may depend on the data with swapped knockoffs.
Above $\hat{\psi}_j$ and $\hat{\psi}^0_j$ are the $j$-th components of $\hat{\psi}$ and $\hat{\psi}^0$.
\end{definition}

 We pause for a moment to explain Definition~\ref{def:knockoff-invariant-function} in generality.
 The intuition is that the partition function $\hat{\psi}$ can only be learnt by looking at a modified data set in which the identities of the real variables are masked by random swapping with their corresponding knockoffs.
Although this (deliberately) introduces some noise, such constraint does not prevent the analyst from learning how to partition the covariate space into useful subgroups because the masked data in $[\mathbf{X},\tilde{\mathbf{X}}]_{\mathrm{swap}(\mathbf{V})}$ still carries valuable information. 
In fact, each column of this matrix contains approximately $n/2$ real observations, and any machine learning algorithm $\smash{\hat{\psi}^0}$ may be utilized to extract useful knowledge from those data.
Further, Definition~\ref{def:knockoff-invariant-function} is highly versatile because it allows one to employ any function $\smash{\hat{\psi}^0}$, which may involve sophisticated statistical or machine learning algorithms, to discover meaningful subgroups.

\subsection{Subgroup learning using interaction-based models} \label{sec:learn-interactions}

An intuitive approach to discovering informative subgroups while satisfying Definition~\ref{def:knockoff-invariant-function} involves implementing the function $\smash{\hat{\psi}^0}$ using parametric regression models with interaction terms.
Suppose that we suspect some of the variables denoted as $X$ may interact with certain covariates $Z$, akin to the earlier illustration in Section~\ref{sec:individualized-hyp}.
For simplicity we assume here that the covariates take binary values, although one could also similarly handle more general categorical covariates, or even continuous-valued covariates, through standard discretization techniques.

In this scenario, the analyst first generates a matrix $\mathbf{V}$ of i.i.d.~Bernoulli random variables, independent of the data.
Then, the analyst proceeds by fitting a sparse generalized linear model (i.e., the lasso) to predict $\mathbf{Y}$ given $\smash{[[\mathbf{X}, \tilde{\mathbf{X}}]_{\mathrm{swap}(\mathbf{V})},\mathbf{Z}]}$, after augmenting the data matrix with all possible pairwise interactions between $\mathbf{Z}$ and $\smash{[\mathbf{X}, \tilde{\mathbf{X}}]_{\mathrm{swap}(\mathbf{V})}}$.
Several algorithms exist for fitting such models \citep{bien2013lasso,lim2015learning,tibshirani2020pliable}, making this approach practical and effective.

After tuning the lasso regularization via cross-validation, let $\hat{\beta}_j$ and $\hat{\beta}_{j+p}$ indicate the estimated main effects for the (randomly swapped) variables $X_j$ and $\smash{\tilde{X}_j}$, for any $j \in [p]$.
Let also $\hat{\gamma}_{l,j}$ and $\hat{\gamma}_{l,j+p}$ indicate the corresponding interaction coefficients involving $Z_l$ and the (randomly swapped) variables $X_j$ and $\tilde{X}_j$, respectively, for all $l \in [m]$ and $j \in [p]$. Given an upper bound $G_{\max}$ on the number of interactions per variable (e.g., $G_{\max}=2$), for all $j \in [p]$ let $\smash{\hat{\mathcal{I}}_j  \subset [m]}$ indicate the subset of $G_{\max}$ covariates with the strongest (nonzero) interactions involving $X_j$ or $\tilde{X}_j$. Above, we understand that any ties are broken at random and $\smash{|\hat{\mathcal{I}}_j| < G_{\max}}$ if the number of covariates with nonzero interaction coefficients is too small.
In other words, we compute
\begin{align*}
  \hat{\mathcal{I}}_j := \big\{ l \in [m] : |\hat{\gamma}_{l,j}|+|\hat{\gamma}_{l,j+p}| > 0, \sum_{l' \neq l} \I{|\hat{\gamma}_{l,j}|+|\hat{\gamma}_{l,j+p}| > |\hat{\gamma}_{l',j}|+|\hat{\gamma}_{l',j+p}|} > m-G_{\max} \big\}.
\end{align*}
This model thus links $X_j$ to at most $G_{\max}$ covariates.
The corresponding functions $\smash{\hat{\psi}_j}$ take values in $\smash{\{1,\ldots,2^{|\hat{\mathcal{I}}_j|}\}}$, which indexes all possible configurations of the covariates in $\hat{\mathcal{I}}_j$. 

Table~\ref{table:interactions-example} helps visualize this approach by highlighting a thought experiment involving 3 explanatory variables and 3 binary covariates.
In this toy example, we imagine that the lasso selects $Z_1, Z_2$ as potentially interacting with $X_1$, and $Z_3$ for $X_2$. 
No interactions with $X_3$ are detected. 
Thus, the population is partitioned into 4 subgroups for $X_1$, 2 subgroups for $X_2$, and 1 trivial subgroup for $X_3$, yielding a total of 7 data-driven hypotheses.


\begin{table}[!htb]
  \caption{Subgroups selected by fitting a lasso model with interactions, and corresponding interpretations of the possible discoveries obtained with our method, in an imaginary toy example with 3 variables and 3 binary covariates. The term ``influences'' is utilized loosely here to indicate a (possibly non-causal) significant conditional association.} \vspace{0.5em}
  \label{table:interactions-example}
    \centering
    \begin{tabular}{ccccc}
        \toprule
      \multirow{2}{*}{Variable} & \multirow{2}{*}{Covariates} & \multicolumn{2}{c}{Partition} & \multirow{2}{4cm}{\centering Interpretation of the findings (upon rejection)}\\
      \cmidrule(l{2pt}r{2pt}){3-4} 
       &  & Label & Definition &  \\
      \midrule
      \multirow{4}{*}{$X_1$} & \multirow{4}{*}{\{$Z_1,Z_2$\}} & 1 & $Z_1 = 0, Z_2 = 0$ & $X_1$ influences $Y$ if $Z_1=0$ and $Z_2=0$\\
       &  & 2 & $Z_1 = 0, Z_2 = 1$  & $X_1$ influences $Y$ if $Z_1=0$ and $Z_2=1$\\
       &  & 3 & $Z_1 = 1, Z_2 = 0$  & $X_1$ influences $Y$ if $Z_1=1$ and $Z_2=0$ \\
       &  & 4 & $Z_1 = 1, Z_2 = 1$  & $X_1$ influences $Y$ if $Z_1=1$ and $Z_2=1$\\
      \hline
      \multirow{2}{*}{$X_2$} & \multirow{2}{*}{\{$Z_3$\}} & 1 & $Z_3 = 0$  & $X_2$ influences $Y$ if $Z_3=0$\\
       &  & 2 & $Z_3 = 1$  & $X_2$ influences $Y$ if $Z_3=1$ \\[0.25em]
      \hline
      $X_3$ & $\emptyset$ & 1 & All individuals  & $X_3$ influences $Y$ \\
        \bottomrule
    \end{tabular}
\end{table}

Note that the parameter $G_{\max}$ generally controls an important trade-off between the power of our method and the interpretability of the findings. A larger value of $G_{\max}$ tends to lead to more specific hypotheses corresponding to smaller subgroups, but it also makes it more difficult to reject those hypotheses, as it will become clear soon.
In fact, partitioning roughly reduces the effective number of samples available during the testing phase by a factor of up to $2^{G_{\max}}$, in the case of binary covariates.
Concretely, all demonstrations presented in this paper will utilize $G_{\max} \in \{1,2\}$, but the optimal value of $G_{\max}$ may generally depend on the data.


\subsection{The subgroup-selective knockoff filter} \label{sec:skf}

For any knockoff-invariant partition function ${\hat{\psi}}$ and any variable index $j \in [p]$, let ${\hat{G}}_j([\mathbf{X}, \tilde{\mathbf{X}}], \mathbf{Y}, \mathbf{Z}) = \max_{z \in \mathbb{R}^m} \hat{\psi}_j(z) \in \mathbb{N}$ be the number of disjoint subgroups induced by $\smash{\hat{\psi}_j}$.
Further, let $\smash{\hat{G} = \sum_{j=1}^{p} {\hat{G}}_j }$. The dependence of $\smash{\hat{\psi}(x)}$, $\smash{\hat{G}}_j$, and $\smash{\hat{G}}$ on the data will not be shown explicitly hereafter unless needed to avoid ambiguity.
Let $\smash{[\mathbf{T},\tilde{\mathbf{T}}] \in \mathbb{R}^{2{\hat{G}}}}$ denote a vector of importance measures for all variables and knockoffs in each region of the covariate space determined by $\hat{\psi}$. 
We will explain later how to compute these importance measures.
For now, note that $\smash{\mathbf{T}}$ (resp.~$\smash{\tilde{\mathbf{T}}}$) is the concatenation of $p$ sub-vectors $\smash{T_j^{g}}$ (resp.~$\smash{\tilde{T}_j^{g}}$) for all $j \in [p]$ and $g \in [{\hat{G}}_j]$, whose elements quantify the importance of $X_j$ (resp.~${\tilde{X}_j}$) in predicting $Y$ within subgroup $g$.

Without loss of generality, the importance measures ${[\mathbf{T},\tilde{\mathbf{T}}]}$ can be written as the output of a (possibly randomized) function $\boldsymbol{\tau}$ applied to a knockoff-augmented data set $[\mathbf{X}, \tilde{\mathbf{X}}],\mathbf{Y},\mathbf{Z}$; i.e.,
\begin{align} \label{eq:def-tau}
  & [\mathbf{T},\tilde{\mathbf{T}}] = \boldsymbol{\tau} \big( [\mathbf{X}, \tilde{\mathbf{X}}],\mathbf{Y},\mathbf{Z} \big)
    = \big[ \bm{t} \big( [\mathbf{X}, \tilde{\mathbf{X}}],\mathbf{Y},\mathbf{Z} \big), \tilde{\bm{t}} \big( [\mathbf{X}, \tilde{\mathbf{X}}],\mathbf{Y},\mathbf{Z} \big) \big].
\end{align}
Above, $\bm{t}$ (resp.~$\tilde{\bm{t}}$) are defined in terms of $\boldsymbol{\tau}$, as its first (resp.~last) ${\hat{G}}$ elements.
In analogy with the familiar case of the model-X knockoff filter, however, achieving false discovery rate control requires imposing some constraints on the function $\boldsymbol{\tau}$.

For any $\mathcal{S} \subseteq [\smash{\hat{G}}]$, whose elements uniquely identify pairs $(j,g)$ for $j \in [p]$ and $\smash{ g \in [{\hat{G}}_j]}$, let $\smash{[\mathbf{X}, \tilde{\mathbf{X}}]_{\mathrm{swap}(\mathcal{S})}}$ be the matrix obtained from $\smash{[\mathbf{X}, \tilde{\mathbf{X}}]}$ after swapping the sub-column $\smash{\mathbf{X}^{(g)}_j}$, which contains all observations of $X_j$ in subgroup $g$, with the corresponding knockoffs, for all $(j,g) \in \mathcal{S}$.
Note the change of notation compared to~\eqref{eq:knock_cond_1}, where swapping was simultaneous for an entire column.
Then, we ask that swapping a real variable with its knockoff within any subgroup should have the only effect of swapping the corresponding importance measures in that subgroup:
\begin{align} \label{eq:swap-tau}
  \boldsymbol{\tau}\big( \mathbf{Y}, [\mathbf{X}, \tilde{\mathbf{X}}]_{\mathrm{swap}(\mathcal{S})},\mathbf{Z} \big)
  =
  \big[
  \bm{t}\big( \mathbf{Y}, [\mathbf{X}, \tilde{\mathbf{X}}],\mathbf{Z} \big),
  \tilde{\bm{t}}\big( \mathbf{Y}, [\mathbf{X}, \tilde{\mathbf{X}}] ,\mathbf{Z}\big) \big]_{\mathrm{swap}(\mathcal{S})}.
\end{align}
In truth, it would be sufficient for this invariance to hold in distribution if the function $\boldsymbol{\tau}$ contained additional independent randomization (e.g., cross-validation to tune some hyper-parameters). However, we can safely ignore this detail here to avoid complicating the notation.


For our method to be powerful, the function $\bm{\tau}$ should be such that a larger value in position $(j,g)$ of $\mathbf{T}$ indicates evidence that $Z_j$ is associated with $Y$ in subgroup $g$. By contrast, a larger value in position $(j,g)$ of $\smash{\tilde{\mathbf{T}}}$ should point to an association of $\smash{\tilde{Z}_j}$ with $Y$, which we know must be spurious.
An example of a valid $\bm{\tau}$ based on the (generalized) lasso is described in Section~\ref{sec:powerful-statistics}. 
That approach is inspired by \citet{li2021searching}  and it is both intuitive and relatively inexpensive, as all of its components can be computed in parallel, but it is not the only possible one.


In general, given any importance measures $[\mathbf{T},\tilde{\mathbf{T}}]$ obtained as described above, we assemble test statistics $\smash{ \mathbf{W} \in \mathbb{R}^{{\hat{G}}}}$ for all selected hypotheses as usual, by computing $\smash{ W^{g}_j = T^{g}_j - \tilde{T}^{g}_j }$ for each pair $(j,g)$.
Finally, we vectorize $\smash{ \mathbf{W}}$ and apply the selective SeqStep+ procedure of \citet{barber2015controlling} (i.e., the final component of the standard knockoff filter) in order to compute an adaptive significance threshold that determines which hypotheses should be rejected.
Figure~\ref{fig:diagram} in Appendix~\ref{app:methods-skf} provides a schematic visualization of this procedure.

The following theorem establishes that our method controls the false discovery rate.
It is worth emphasizing that this result is far from trivial and does not stem directly from \citet{candes2018} or \citet{li2021searching}. While our proof strategy is inspired by those earlier works, our scenario is inherently more complex. 
In fact, our hypotheses are not predetermined but instead automatically selected using the adaptive partition function $\hat{\psi}$, which depends on the same data used to compute the test statistics.
This is a delicate situation that can generally introduce selection bias \citep{assmann2000subgroup}, as previewed earlier in Figure~\ref{fig:experiment-heterogeneous-1}.
Nevertheless, our method is meticulously designed to circumvent such selection bias through the incorporation of independent randomization within the matrix $\mathbf{V}$, which is used by the analyst to partially mask the training data for $\hat{\psi}$.

\vspace{0.5em}
\begin{theorem} \label{thm:coin-flip}
Consider a knockoff-augmented data set $\smash{[\mathbf{X}, \tilde{\mathbf{X}}],\mathbf{Y},\mathbf{Z}}$, and let $\mathbf{V} \in \{0,1\}^{n \times p}$ be i.i.d.~Bernoulli random variables generated by the analyst, independent of the data.
Let $\hat{\psi}$ be a knockoff-invariant partition function with total size $\smash{\hat{G} = \sum_{j=1}^{p} \hat{G}}_j$, trained looking only at $\smash{[\mathbf{X}, \tilde{\mathbf{X}}]_{\mathrm{swap}(\mathbf{V})},\mathbf{Y},\mathbf{Z}}$.
Let also $\smash{\mathbf{W} \in \mathbb{R}^{{\hat{G}}}}$ indicate test statistics, for the hypotheses defined by $\hat{\psi}$, computed as described above using a function $\boldsymbol{\tau}$ that satisfies~\eqref{eq:swap-tau}.
Suppose $\smash{\mathbf{U} \in \{\pm 1\}^{\hat{G}}}$ is a random vector with independent entries such that: $\smash{U^{g}_{j}= \pm 1}$ with probability $1/2$ if $\smash{\mathcal{H}^{(g)}_{0,j}}$ in~\eqref{eq:null-hyp-env} is true and ${U^{g}_j = +1}$ otherwise, for all $j \in [p]$ and ${g \in [\hat{G}_j]}$. 
Then,  $\smash{ \mathbf{W} \mid \hat{\psi}\; \oset{d}{=}\; \mathbf{W} \odot \mathbf{U}  \mid \hat{\psi} }$.
\end{theorem}
\vspace{0.5em}

The proof of Theorem~\ref{thm:coin-flip} is in Appendix~\ref{app:proofs}.
An immediate corollary is that the standard knockoff filter applied to statistics $\mathbf{W}$ computed as described above will control the false discovery rate for~\eqref{eq:null-hyp-env} conditional on $\smash{\hat{\psi}}$; e.g., see Section~\ref{sec:knockoffs-review} and \citet{candes2018}.

\subsection{Computing powerful test statistics for the subgroup-selective knockoff filter} \label{sec:powerful-statistics}

We describe here an implementation of our method based on generalized linear models.
As the property in~\eqref{eq:swap-tau} requires swapping $X_j$ and $\tilde{X}_j$ in subgroup $g$ to have the only effect of swapping the corresponding $\smash{T_j^{g}}$ and $\smash{\tilde{T}_j^{g}}$, an intuitive approach inspired by~\cite{li2021searching} is the following.
First, a sparse generalized linear model (e.g., the lasso) is trained to predict $\mathbf{Y}$ given $\smash{[[\mathbf{X}, \tilde{\mathbf{X}}]_{\mathrm{swap}(\mathbf{V})},\mathbf{Z}]}$, tuning the regularization via cross-validation.
The absolute values of the regression coefficients serve as ``prior'' importance measures, $\smash{T_j^{\text{prior}}}$ and $\smash{\tilde{T}_j^{\text{prior}}}$, for all variables and knockoffs indexed by $j \in [p]$.
These are combined pairwise into a weight $\pi_j$ for each $j$; for example, as $\smash{ \pi_j = \zeta(T^{\text{prior}}_j+\tilde{T}^{\text{prior}}_j) }$, where $\zeta$ is a positive and decreasing function such as $\zeta(t) = 1/(0.05+t)$. Larger values of $\pi_j$ suggest the $j$-th variable is more likely to have a significant effect among some individuals. Similar weights can also be defined for the covariates.
Then, separately for each $(j,g)$, a new model is fitted using only the data from the $g$-th subgroup to predict $\mathbf{Y}^{(g)}$ given $\smash{[ [\mathbf{X}^{(g)}, \tilde{\mathbf{X}}^{(g)}]_{\mathrm{swap}(\mathbf{V})}, \mathbf{Z}^{(g)}]}$, after restoring the true variable and knockoff identities in the $j$-th and $(j+p)$-th columns. See Figure~\ref{fig:diagram_2} for a schematic visualization of this algorithm.

This model utilizes feature-specific regularization that depends on two hyper-parameters, $\lambda^{(j,g)}>0$ and $\xi^{(j,g)} \in [0,1]$, both tuned by cross-validation, and on $\pi$. Specifically, the penalty for the $l$-th variable is $\smash{\lambda^{(j,g)}_l = \lambda^{(j,g)} (1-\xi^{(j,g)}) + \xi^{(j,g)} \pi_l}$, for all $l \in [p]$.
If $\smash{\xi^{(j,g)} = 0}$, this reduces to a standard lasso looking only at subgroup $g$. But, in general, larger values of $\xi^{(j,g)}$ can make our approach more powerful, as we gather strength from the data in all subgroups. In fact, null variables will tend to receive smaller values of $\pi$ and will thus be less likely to be incorrectly selected by the final model, thereby reducing the noise in the test statistics.
Of course, the weights $\pi$ may not always be informative, hence why $\xi^{(j,g)}$ is tuned by cross-validation. For that purpose, an expensive two-dimensional grid search can be avoided by tuning first $\lambda^{(j,g)}>0$ and then $\xi^{(j,g)}$.

Finally, the importance measures for the $j$-th variable and knockoff in subgroup $g$ are defined as the absolute values of the regression coefficients for $X_j$ and $\tilde{X}_j$, respectively.
It is easy to prove this procedure satisfies~\eqref{eq:swap-tau}. In fact, conditional on $\smash{\mathbf{Y}, [\mathbf{X},\tilde{\mathbf{X}}]_{\mathrm{swap}(\mathbf{V})}, \mathbf{Z}}$, swapping all observations of $X_j$ in subgroup $g$ with the corresponding $\smash{\tilde{X}_j}$ for any pair $(j,g)$ simply results in swapping $\smash{T_j^{(g)}}$ with $\smash{\tilde{T}_j^{(g)}}$, as $\pi$ is unperturbed because it only depends on $\smash{[\mathbf{X},\tilde{\mathbf{X}}]_{\mathrm{swap}(\mathbf{V})}}$.

While it is convenient to think about the lasso, the main idea behind our approach is applicable with any model. The key to achieving~\eqref{eq:swap-tau} is that the importance measures indexed by $(j,g)$ are computed by a model that sees the other variables and the observations in other subgroups only through the lenses of the data with randomly swapped knockoffs.
The weights $\pi$ must be estimated based on $\smash{[[\mathbf{X}, \tilde{\mathbf{X}}]_{\mathrm{swap}(\mathbf{V})},\mathbf{Z}]}$ to make the test statistics mutually independent, which is required for false discovery rate control; see the proof of Theorem~\ref{thm:coin-flip}.
Further, revealing the identities of $X_j$ and $\tilde{X}_j$ within subgroup $g$ is essential to achieve non-trivial power---otherwise, it would be impossible to tell important variables apart from knockoffs.
A generalization of the above solution would involve estimating different $\pi$ weights in different subgroups, as sketched by Figure~\ref{fig:diagram_2} in Appendix~\ref{app:methods-skf}. That approach still satisfies~\eqref{eq:swap-tau} if the models utilized to compute the priors only look at the randomly swapped data in $[\mathbf{X}, \tilde{\mathbf{X}}]_{\mathrm{swap}(\mathbf{V})}$.

\section{Empirical demonstrations}  \label{sec:empirical}

\subsection{Demonstrations with synthetic data} \label{sec:experiments}

We generate synthetic data with 20 binary variables and 80 covariates.
The first 20 covariates are independently sampled from a Bernoulli(0.5) distribution.
The remaining 60 covariates follow a Gaussian autoregressive model of order one with correlation parameter 0.5.
The outcome is generated from a linear model with heterogeneous effects and homoscedastic standard Gaussian noise.
The linear coefficients in this model are zero for half of the variables, for all of the binary covariates, and for half of the 60 continuous covariates.
That is, the expected outcome for the $i$-th individual is $\smash{ \E{Y^{i} \mid X^{i}, Z^{i}} = \sum_{j=1}^{p} X^{i}_j \beta_j^{i} + \sum_{j=21}^{m} Z^{i}_j \gamma_j}$, with $\smash{\beta_j^{i} = \bar{\beta}_j Z_{l_{j,1}}^{i} Z_{l_{j,2}}^{i} }$,
for $p = 20$ and $m = 80$.
Above, the $\beta$ coefficients are individual-specific, while the $\gamma$ are constant.
The parameters $\bar{\beta}_j$ and $\gamma_j$ are initialized with absolute value equal to 4 and independent random signs, for each $j \in [p]$.
Then, separately for each variable $j$ and individual $i$, the individualized coefficient $\smash{\beta_j^{i}}$ is set equal to the product of $\smash{\bar{\beta}_j}$ and the two binary covariates $Z_{l_{j,1}}$ and $Z_{l_{j,2}}$, for some $l_{j,1}, l_{j,2} \in [20]$. Thus, each variable $X_j$ has an effect on $Y$ only within the subgroups with $Z_{l_{j,1}}=1$ and $Z_{l_{j,2}}=1$, which contains approximately 1/4 of all individuals if $l_{j,1} \neq l_{j,2}$. The indices  $l_{j,1}, l_{j,2}$ are sampled with replacement from $\{1,\ldots,20\}$, independently for each~$j$.
Our goal is to discover which variables are non-null within which subgroups, as powerfully (seeking more numerous findings) and precisely (seeking more informative subgroups) as possible.

The subgroup-selective knockoff filter is compared to the three benchmarks outlined in Section~\ref{sec:preview}. The first benchmark is the vanilla knockoff filter of~\cite{candes2018}, which controls the false discovery rate but can only test population-wide hypotheses.
The second benchmark is the data-splitting version of our method, which utilizes half of the samples to learn the partition and the other half to test the conditional hypotheses. This tests the same hypotheses and enjoys the same statistical guarantees as the subgroup-selective knockoff filter but is less powerful.
The third benchmark is a greedy version of our method that naively utilizes all data twice, first to select the partitions and then to test the conditional hypotheses, without randomly masking the identity of the knockoffs; this does not control the false discovery rate due to selection bias.

All methods are compared in terms of their average proportion of false discoveries, their power, and the informativeness of their findings.
The latter is quantified by two complementary metrics: the average homogeneity of the rejected hypotheses, defined as the proportion of individuals in the reported subgroup for which the variable of interest has a non-zero coefficient within the data-generating model, and the corresponding heterogeneity of individual effects, defined as the standard deviation of the true model coefficients.

Figure~\ref{fig:experiment-heterogeneous-1}, previewed in Section~\ref{sec:skf}, reports the performances of all methods over 100 independent experiments.
As discussed in Section~\ref{sec:preview}, these results confirm the validity of our approach and demonstrate its advantages relative to the naive, data-splitting, and vanilla benchmarks. In summary, our method leads to discoveries that are more informative than those of the vanilla knockoff filter, more numerous than those of the sample-splitting benchmark, and more reliable than those of the naive greedy benchmark.

Appendix~\ref{app:experiments-discovery} presents additional numerical results, demonstrating the use of the subgroup-selective knockoff filter in a ``transfer learning'' setting inspired by~\cite{li2022transfer}. The goal there is complementary: we wish to detect variables with a robust association across different subgroups. This can be achieved with an extension of our method described in Appendix~\ref{app:methods-rskf}, which tests partial conjunctions of the hypotheses in~\eqref{eq:null-hyp-env}.

\subsection{Demonstrations with data from a randomized experiment} \label{sec:data}

We conduct numerical experiments based on a data set from a 2-week long randomized experiment involving 80,000 blood donors in China \citep{sun2019mobile}, which was designed to investigate the effectiveness of different donation incentives.
In this experiment, 80,000 individuals were randomly divided into 7 groups: a control group of size 14,000 and 6 treated groups of size 11,000 each.
The treated groups received a text message with a reminder to donate, while the control group received nothing.
Treated groups 2--5 received further encouragement to donate: the promise of an individual monetary reward (equivalent to \$5.00--\$8.30, depending on the amount of blood donated) for group 2; a suggestion to bring a friend for group 3, both the individual monetary reward and the friend suggestion for group 4; a suggestion to bring a friend and the promise of a group reward (equivalent to the individual reward) for group 5; a suggestion to bring a friend, the promise of a group reward (as for the previous group), and the promise of a small extra gift for group 6.
See Table~\ref{table:data-blood-groups} in Appendix~\ref{app:data-description} for a summary of this experimental design.
In the end, 797 individuals made a donation.
These data also contain covariates including age, sex, weight, blood type, marital status, education level, occupation, residency status (local or non-local), and time since the last blood donation.
Missing values (less than 4\%) are imputed with the corresponding sample median (or mode, if categorical).

To simplify the task of generating knockoffs, we approximate the joint distribution of the binary treatments $X_1,\ldots,X_5$ by imagining that different individuals were assigned to one of the six possible groups listed in Table~\ref{table:data-blood-groups} independently of one another, with probability proportional to the observed group size. This model is not perfect because the treatment group sizes were in truth fixed a priori, but the simplification is useful as it leads to a manageable joint treatment distribution for which exact knockoffs can be generated by the Metropolized algorithm of~\cite{bates2020metropolized}.
Goodness-of-fit diagnostics for the knockoffs thus obtained are reported in Tables~\ref{tab:knockoffs-1}--\ref{tab:knockoffs-2}.
In principle, it would also be possible to generate knockoffs without the above independence assumption using an algorithm similar to that of \citet{sesia2021false}, but that would be more technically involved and seems likely unnecessary in this case.

Given that the positive outcomes from this randomized experiment were very rare (approximately 1\%), we expect that any conditional testing method may not be able to make many discoveries with these data.
Therefore, we find it more interesting to conduct a semi-synthetic analysis in which the true donation events are replaced by simulated outcomes from an imaginary causal model conditional on the treatments and covariates. 
This setup remains quite realistic, as it involves real variables, but it reduces the outcome imbalance and increases the effective sample size. 
Consequently, the analysis will be more informative regarding the effectiveness of the our methodology applied to large data sets. 
Further, as the ground truth is known exactly in such a controlled setting, we can have access to useful diagnostics, including the type-I errors.

For each of the 80,000 individuals, an imaginary donation outcome is simulated from a logistic model based on the $m$ real covariates $Z \in \mathbb{R}^m$ and the treatment variables $X \in \mathbb{R}^5$; i.e., $$\smash{ \text{logit}\left(\P{Y = 1 \mid X, Z} \right)  = -c + \sum_{j=1}^{m} Z_j b_j + a \sum_{j=1}^{5} X_j \cdot g_j(Z) },$$
where $c \in \mathbb{R}$ is an intercept, $b_1, \ldots, b_m \in \mathbb{R}$ are linear coefficients, and each $g_j$ is a binary-valued non-linear function of $Z$;
see Appendix~\ref{app:analysis-semi-synthetic} for more details.
The first treatment (text reminder) is effective for individuals who did not donate recently and for those who are non-residents. The individual reward is effective for students, and twice as much for those who are male. The friends request is effective for female or unmarried individuals. The group reward is effective for males or students. The small gift is effective for individuals with fewer than 16 years of education. The goal is to discover which treatments are effective within which subgroups, as powerfully and precisely as possible.
The subgroup-selective knockoff filter is applied as in the previous section, defining the subgroup-specific hypotheses based on the top two candidate covariates selected by a (logistic) lasso model fitted on the data with randomly swapped knockoffs.

\begin{figure}[!htb]
  \centering
  \includegraphics[width=0.95\textwidth]{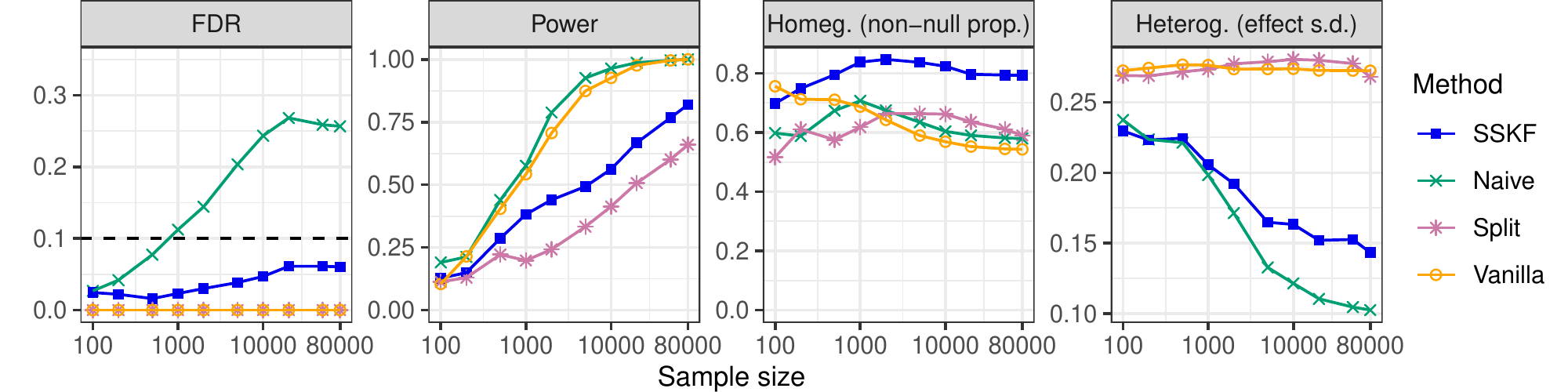}
  \caption{Performance of the subgroup-selective knockoff filter and benchmarks in simulations based on data from a randomized blood donation experiment. Other details are as in Figure~\ref{fig:experiment-heterogeneous-1}.}
  \label{fig:experiment-blood}
\end{figure}

Figure~\ref{fig:experiment-blood} summarizes the results of 100 experiments with independent knockoffs and outcomes, varying the sample size. 
Each time, a random subset of individuals is analyzed.
The performance of the subgroup-selective knockoff filter is quantified in terms of false discovery rate and power, as well as in terms of the homogeneity of the selected hypotheses, as in Section~\ref{sec:experiments}. The same benchmarks as in Section~\ref{sec:experiments} are also considered here.
The subgroup-selective knockoff filter controls the false discovery rate and is more powerful than data splitting. 
As expected, the naive approach does not control the false discovery rate, while the vanilla knockoff filter can only discover less informative population-wide associations.
The list of discoveries obtained in the first experiment with sample size 80,000 is provided in Table~\ref{table:experiment-blood-disc}.
Among the 11 findings, that of the friends request among unmarried females has the largest test statistic; indeed, we know this treatment has a causal effect for all those individuals. Although no type-I errors are made in this case, not all findings are equally informative. For example, the reminder can be effective among both male and female residents, but the subgroup-selective knockoff filter does not tell us that it is only causal for the 60\% of them who are married.
In fact, this interaction is not discovered because the initial model fitted on the data with randomly swapped knockoffs failed to select the corresponding covariate, likely due to the limited sample size.

\section{Discussion} \label{sec:discussion}

This paper has extended the model-X framework of~\cite{candes2018} to enable more informative inferences in the presence of sample heterogeneity and interactions.
Flexibility is the main strength of the proposed method, which can borrow strength from any statistical or machine learning model while providing rigorous guarantees in finite samples.

A promising opportunity for future research is to apply the subgroup-selective knockoff filter to large-scale data from genome-wide association studies \citep{storey2003statistical}, such as the UK Biobank \citep{bycroft2018} or the Millions Veteran Project \citep{gaziano2016million}.
The populations sampled by those studies are heterogeneous \citep{peterson2019genome}, and our method may be helpful to discover possible gene-environment interactions \citep{moore2006flexible}. Further, the abundance of variables and observations in those studies will translate into further advantages for our methodology, which generally tends to perform better when applied to larger-scale data sets. 
Computing powerful test statistics based on extremely big data sets involves some computational challenges, but the model-X framework has already been demonstrated to be quite scalable \citep{liu2020fast,sesia2020multi,sesia2021false}.

Further, it would be interesting to explore whether our inferences could be (at least partly) derandomized using techniques inspired by  \citet{ren2022derandomized} and \citet{ren2023derandomizing}.

A software implementation of our methods is available online at~\url{https://github.com/msesia/i-modelx}, along with the code needed to reproduce our numerical results.

\section*{Acknowledgements} \label{sec:ack}

M.S.~is also affiliated with the Department of Computer Science at the University of Southern California.
M.~S.~was supported by NSF grant DMS 2210637 and an Amazon Research Award.
The authors are grateful to Edgar Dobriban, Yingying Fan, Nikolaos Ignatiadis, and Stefan Wager for providing insightful feedback about an earlier version of this manuscript.


\printbibliography

\appendix

\renewcommand{\thesection}{A\arabic{section}}
\renewcommand{\theequation}{A\arabic{equation}}
\renewcommand{\thetheorem}{A\arabic{theorem}}
\renewcommand{\theproposition}{A\arabic{proposition}}
\renewcommand{\thelemma}{A\arabic{lemma}}
\renewcommand{\thetable}{A\arabic{table}}
\renewcommand{\thefigure}{A\arabic{figure}}
\renewcommand{\thealgocf}{A\arabic{algocf}}
\setcounter{figure}{0}
\setcounter{table}{0}
\setcounter{proposition}{0}
\setcounter{theorem}{0}
\setcounter{lemma}{0}

\section{Background on population-wide conditional testing} \label{app:standard-hypotheses}

\subsection{The interpretation of conditional independence hypotheses} 

The model-X problem studied by \citet{candes2018} can be thought of as testing, for all $j \in [p]$, whether $X_j$ is associated with $Y$ given $Z$ and all variables excluding $X_j$ (i.e., $X_{-j}$); that is, whether the null hypothesis defined in~\eqref{eq:null-hyp} holds true: $\mathcal{H}_{0,j} : Y \indep X_j \mid X_{-j}, Z$.
For example, if one assumed a generalized linear model for $Y \mid X,Z$ (which we do not do here), then the hypothesis in~\eqref{eq:null-hyp} would reduce, under relatively mild assumptions on $P_{X \mid Z}$,  to stating that the linear coefficient for $X_j$ is zero~\citep{candes2018}.

More generally, without any parametric model for the distribution of the outcome, a rejection of~\eqref{eq:null-hyp} can be interpreted as a discovery that $X_j$ has some effect on $Y$ conditional on the other observed variables, for at least some of the individuals in the population. (Note that we are not assuming the individuals are identically distributed.)
These discoveries can be generally useful to screen variables in high-dimensional data analyses \citep{candes2018}, to help prioritize follow-up studies \citep{sesia2020multi}, and in some cases even to make approximate causal inferences \citep{li2021searching}.
In fact, under some additional assumptions such as the absence of unmeasured confounders, a rejection of~\eqref{eq:null-hyp} can sometimes be rigorously interpreted as stating that $X_j$ causes $Y$ \citep{bates2020}. 
See Appendix~\ref{app:causal-inference} and Figure~\ref{fig:dag} for details on the connections to causal inference.

\subsection{The connection to causal inference} \label{app:causal-inference}

\begin{figure}[H]
  \centering
  \includegraphics[]{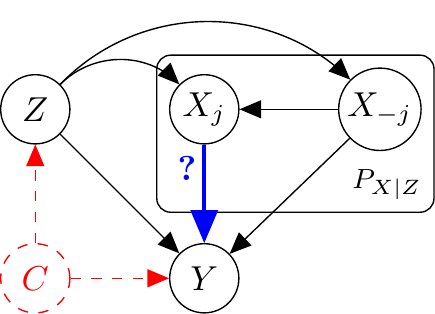}
  \caption{Graphical representation of a non-parametric causal model linking the treatment ($X$), the outcome ($Y$), the measured covariates ($Z$), and possibly also other  unmeasured covariate ($C$). Our goal is to test whether a particular treatment $X_j$ has any causal effect on the outcome. The joint distribution of $X \mid Z, C$ is assumed to be known and may depend only on $Z$, so that $X \indep C \mid Z$.
}
  \label{fig:dag}
\end{figure}

In addition to the basic setup of Section~\ref{sec:notation}, imagine the variables $X$ are randomized treatments that may depend on the observed covariates $Z$ but are independent of anything else.
Further, suppose the data distribution takes the form of a structural equation model~\cite{bollen2014structural, pearl_2009} in which $Z$ and $X$ may cause $Y$, but not the other way around, as visualized by the directed acyclic graph in Figure~\ref{fig:dag}.
It is easy to see this setup implicitly rules out confounding in a test of conditional independence between $X$ and $Z$, because the treatments are conditionally independent of any other unmeasured covariate $C$.
Therefore, to test whether $X_j$ has a causal effect on $Y$, it suffices to test whether the conditional independence hypothesis in~\eqref{eq:null-hyp}.
\begin{proposition}[From~\cite{bates2020}] \label{prop:causal-reduction}
Let $C$ be any unmeasured covariate. If $X \indep C \mid Z$, a valid test of the conditional independence null hypothesis $\mathcal{H}_{0,j}$ in \eqref{eq:null-hyp} is also a valid test of the stronger null hypothesis
\begin{align} \label{eq:causal-hyp}
  \mathcal{H}_{0,j}^* : Y \indep X_j \mid Z, X_{-j}, C.
\end{align}
\end{proposition}
\begin{proof}
The proof is analogous to that of Proposition~1 in~\cite{bates2020}.
Suppose $\mathcal{H}_{0,j}^* : Y \indep X_j \mid Z, X_{-j},C$ is true.
Then, it follows from $C \indep X_{j} \mid Z, X_{-j}$ that $(Y,C) \indep X_{j} \mid Z, X_{-j}$. Therefore, $\mathcal{H}_{0,j}$ must also be true.
\end{proof}

\section{Method schematics} \label{app:methods-skf}

\begin{sidewaysfigure}
  \centering
  \includegraphics[width=20cm]{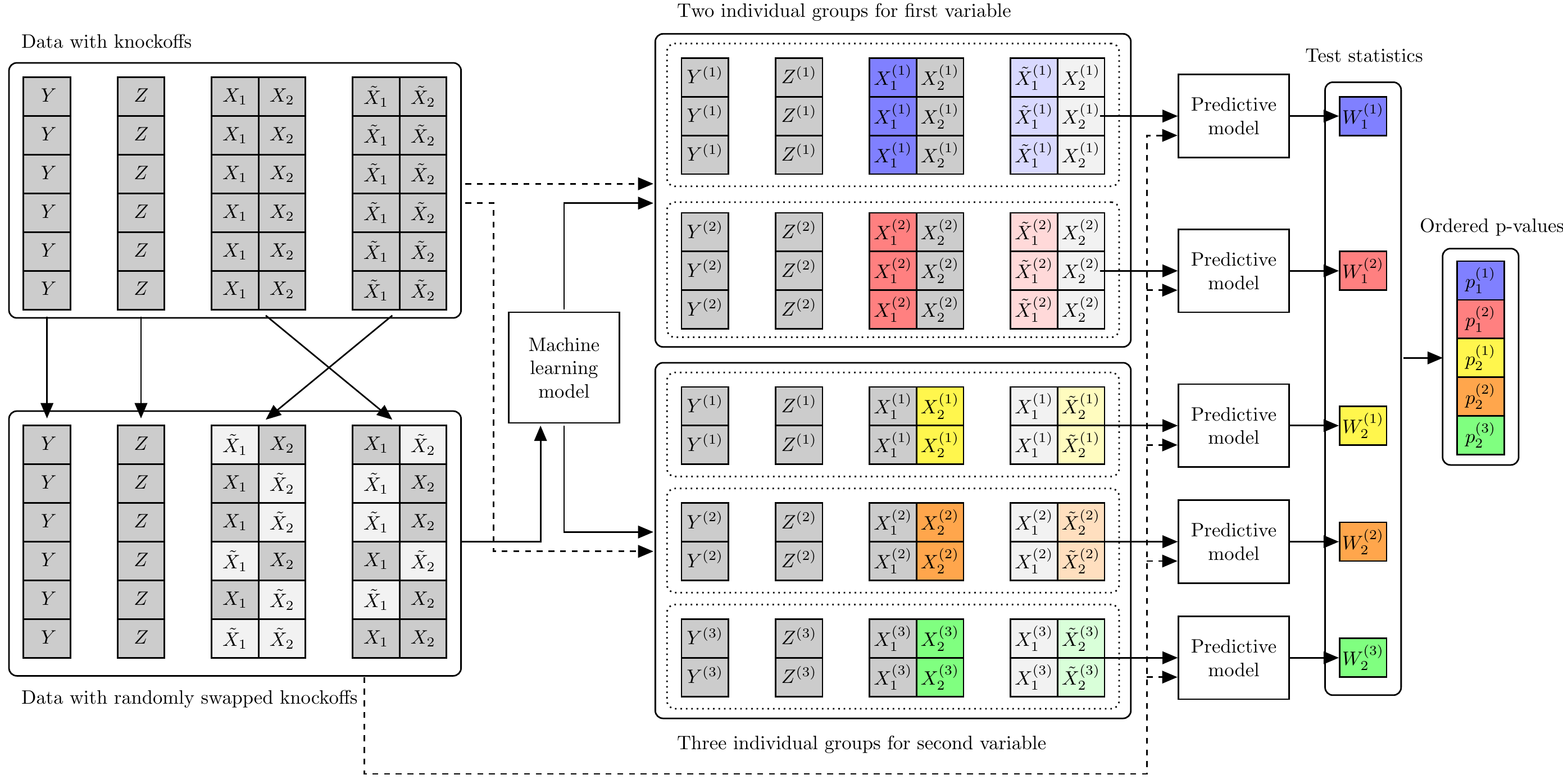}
  \caption{Schematic of the subgroup-selective knockoff filter for subgroup-specific conditional testing. The statistics for distinct combinations of variables and individual groups are shown in different colors.}
  \label{fig:diagram}
\end{sidewaysfigure}

\begin{sidewaysfigure}
  \centering
  \includegraphics[width=23cm]{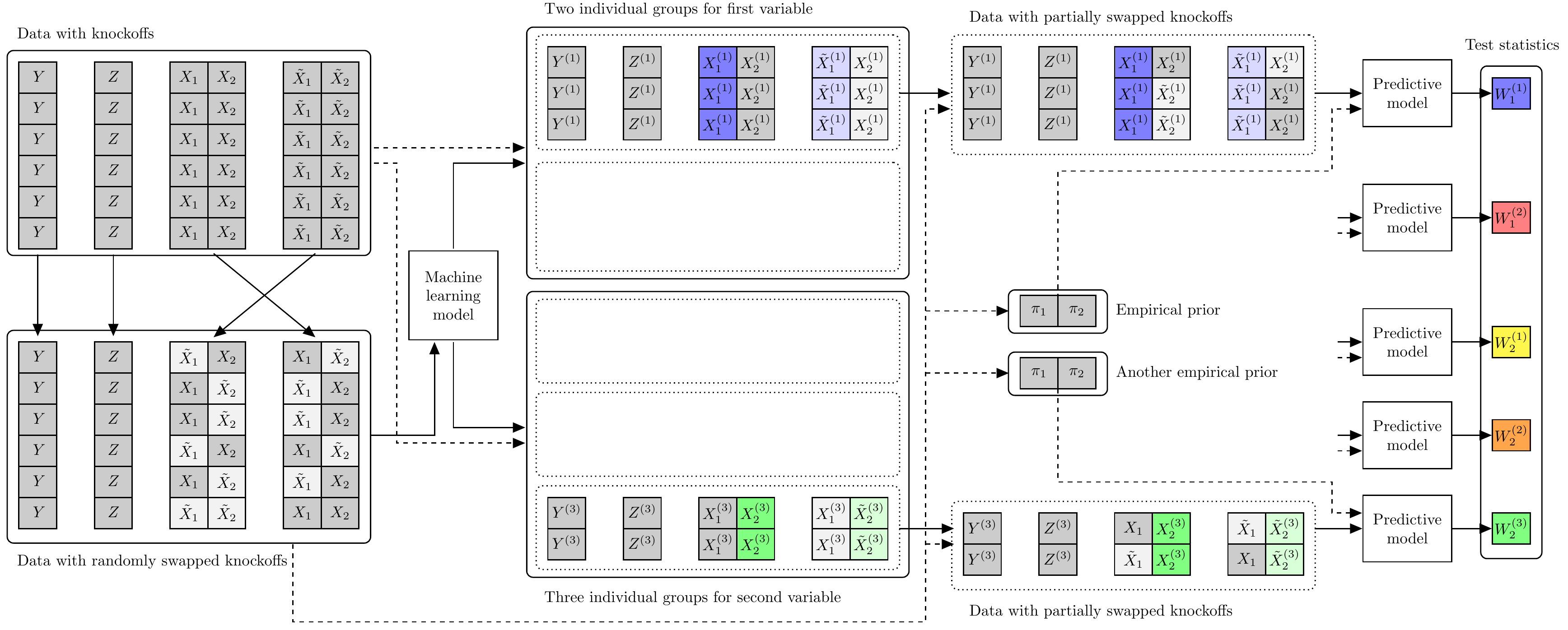}
  \caption{Schematic of the subgroup-selective knockoff filter for subgroup-specific conditional testing, focusing on the computation of the statistics for the first variable in subgroup 1 (blue) and for the second variable in subgroup 3 (green). The test statistics are based on a simple empirical prior for the two variables that is learnt from the full data set with randomly swapped knockoffs. Other details are as in Figure~\ref{fig:diagram}.}
  \label{fig:diagram_2}
\end{sidewaysfigure}




\FloatBarrier
\section{The robust subgroup-selective knockoff filter}  \label{app:methods-rskf}

The subgroup-selective knockoff filter was developed to test subgroup-specific hypotheses~\eqref{eq:null-hyp-env} of the type
\begin{align*}
  \mathcal{H}^{(g)}_{0,j} : Y^{(g)} \indep X^{(g)}_j \mid X^{(g)}_{-j}, Z^{(g)},
\end{align*}
for all variables $j$ and subgroups $g$, conditional on a random partition $\hat{\psi}$ based on the data with randomly swapped knockoffs.
Here, we present a further methodological extension that makes it possible to test the following {\em partial conjunction}~\cite{friston2005conjunction,benjamini2008screening} hypotheses:
\begin{align} \label{eq:null-pici}
\mathcal{H}^{\mathrm{pc},r}_{0,j} : \left| \left\{ g \in [G] : \mathcal{H}^{(g)}_{0,j} \text{ is true} \right\} \right| > G-r,
\end{align}
for any fixed $r \in \{1,\ldots, G\}$. 
This is inspired by~\citet{li2021searching}.
Intuitively, the partial conjunction hypothesis states that the number of subgroups for variable $j$ in which $\mathcal{H}^{(g)}_{0,j}$~\eqref{eq:null-hyp-env} is false is strictly smaller than $r$.
Therefore, rejecting this null suggests $X_j$ is associated with $Y$ in at least $r$ distinct subgroups.
If $r = 1$, the partial conjunction hypothesis reduces to the original population-wide hypothesis of~\citet{candes2018}, in~\eqref{eq:null-hyp}; in that case, rejecting~$\mathcal{H}^{\mathrm{pc},r}_{0,j}$ suggests the existence of least one subgroup in which $X_j$ is conditionally associated with $Y$.
By contrast, rejecting the hypothesis in~\eqref{eq:null-pici} for $r=G$ requires evidence of a conditional association between $X_j$ and $Y$ across {\em all} subgroups.

One reason why it can be interesting to test $\mathcal{H}^{\mathrm{pc},r}_{0,j}$ is that this hypothesis may point towards the discovery of variables whose association with $Y$ is {\em robust} to changes in the covariate distribution~\citep{li2021searching}.
Intuitively, testing $\mathcal{H}^{\mathrm{pc},r}_{0,j}$ can be seen as a way for discovering conditional associations that do {\em not} involve interactions, thus taking a point of view complementary to that of the subgroup-selective knockoff filter.
The $p$ partial conjunction hypotheses defined in~\eqref{eq:null-pici} are relatively straightforward to test while provably controlling the false discovery rate, if one starts with a collection $\mathbf{W}$ of subgroup-selective knockoff filter statistics satisfying the flip-sign property in Theorem~\ref{thm:coin-flip}: $\mathbf{W} \mid \hat{\psi} \; \oset{d}{=}\; \mathbf{W} \odot \mathbf{U} \mid \hat{\psi}$. In fact, it suffices to apply the multi-environment knockoff filter of~\citet{li2021searching} to these statistics, as explained below.

For each $j \in \{1,\ldots,p\}$, let $n^-_j$ count the negative signs among the $W_{j}^{(g)}$ statistics for all subgroups $g$, and let $\smash{n^0_j}$ be the number of zeros among them. For simplicity, assume $\smash{\hat{\psi}}$ partitions the covariate space into $G$ subgroups for each variable; this is without loss of generality, as it is otherwise sufficient to set the remaining undefined statistics to zero if $\hat{G}_j < G$ for some $j \in [p]$.
Then, compute
\begin{align}  \label{eq:pici-pvalues}
\begin{split}
  p_j^{\mathrm{pc},r}
  & = \Psi\left( n^-_j -1, (G-r+1-n^0_j) \lor 0, \frac{1}{2} \right) + U_j \cdot \Psi'\left( n^-_j, (G-r+1-n^0_j) \lor 0, \frac{1}{2}\right),
\end{split}
\end{align}
where $\Psi(\cdot, m,\pi)$ is the binomial cumulative distribution function, $\Psi'(\cdot,m,\pi)$ is the corresponding probability mass, and $U_j$ is a uniform random variable on $[0,1]$ independent of everything else.
Further, define
\begin{align} \label{eq:pici-order}
  |W_j^{\mathrm{pc},r}| = \bar{w} \left( |W_{j}^{(1)}|, \dots,|W_{j}^{G}| \right),
\end{align}
for some symmetric function $\bar{w}$, such as that which multiplies the top $r$ largest entries by absolute value:
\begin{align*}
w \left( |W_{j}^{(1)}|, \dots,|W_{j}^{(G)}| \right) = \prod_{g = 1}^r  \bar{|W|}_{j}^{(G-g+1)}.
\end{align*}
Above, $\bar{|W|}_{j}^{(g)}$ are the order statistics for $\{|W_{j}^{(1)}|, \dots,|W_{j}^{(G)}|\}$.
Then, the false discovery rate for~\eqref{eq:null-pici} can be controlled by applying the usual selective SeqStep+ sequential testing procedure of~\citet{barber2015controlling}, thanks to a relatively simple extension of Theorem~1 from~\citet{li2021searching}.
\begin{theorem} \label{th:seqstep-fdr}
Selective SeqStep+ applied to the p-values \eqref{eq:pici-pvalues} ordered by~\eqref{eq:pici-order} controls the false discovery rate for~\eqref{eq:null-pici} if the statistics $\mathbf{W}$ satisfy the flip-sign property in Theorem~\ref{thm:coin-flip}: $\mathbf{W} \mid \hat{\psi} \; \oset{d}{=}\; \mathbf{W} \odot \mathbf{U} \mid \hat{\psi}$.
\end{theorem}
\begin{proof}
This result follows immediately from Theorem~1 in~\citet{li2021searching} with a simple extension of Proposition~5 therein.
The original statement of Proposition~5 in~\citet{li2021searching} assumed the statistics $\mathbf{W}$ to be computed using data collected from separate and a-priori fixed experimental settings (or {\em environments}), but it is easy to see that their proof only requires the flip-sign property established by our Theorem~\ref{thm:coin-flip}.
\end{proof}

Alternatively, the false discovery rate for~\eqref{eq:null-pici} could be (approximately) controlled by applying to the p-values \eqref{eq:pici-pvalues} ordered by~\eqref{eq:pici-order} the accumulation test~\citep{li2017accumulation} instead of SeqStep+; this can be easily proved with an analogous extension of Theorem~2 from~\citet{li2021searching}.

\FloatBarrier
\section{Additional results from numerical experiments} \label{app:experiments}

\subsection{Experiments in a ``transfer learning'' setting} \label{app:experiments-discovery}

Here, we investigate the performance of the robust subgroup-selective knockoff filter from Appendix~\ref{app:methods-rskf} applied in a ``transfer learning'' setting inspired by~\cite{li2022transfer}. 
The goal is complementary to that of the experiments in Section~\ref{sec:experiments}: we wish to detect variables with a robust association across subgroups with different covariate distributions.
In other words, we want to find which variables maintain their association with $Y$ within a future data set with covariate shift, in which the distribution of $Z$ differs from the current one but the true model for $Y \mid X,Z$ is the same~\citep{li2022transfer}.
To simulate this scenario, we generate data from a model similar to that in Section~\ref{sec:experiments}, but now only half of the non-null variables interact with the covariates; i.e., $\smash{\beta_j^{(i)}} = \bar{\beta}$ for half of the variables.
Further, the data dimensions are increased to ensure sufficiently many discoveries can be made. Specifically, the number of variables is $p=40$, while the number of covariates is $m=160$, of which 120 are continuous.
The covariate shift is imagined to be such that the 40 binary covariates are always equal to zero instead of following a symmetric Bernoulli distribution.
Therefore, the goal is to identify the $20$ variables whose association does not involve interactions, as the covariate shift would make the others irrelevant.
These data are analyzed as in Section~\ref{sec:experiments}, with the only difference that the subgroup-selective knockoff filter is replaced by its robust version outlined in Appendix~\ref{app:methods-rskf}, which is designed to test partial conjunctions of the subgroup-specific conditional hypotheses~\eqref{eq:null-hyp-env}.

\begin{figure}[H]
  \centering
  \includegraphics[width=\textwidth]{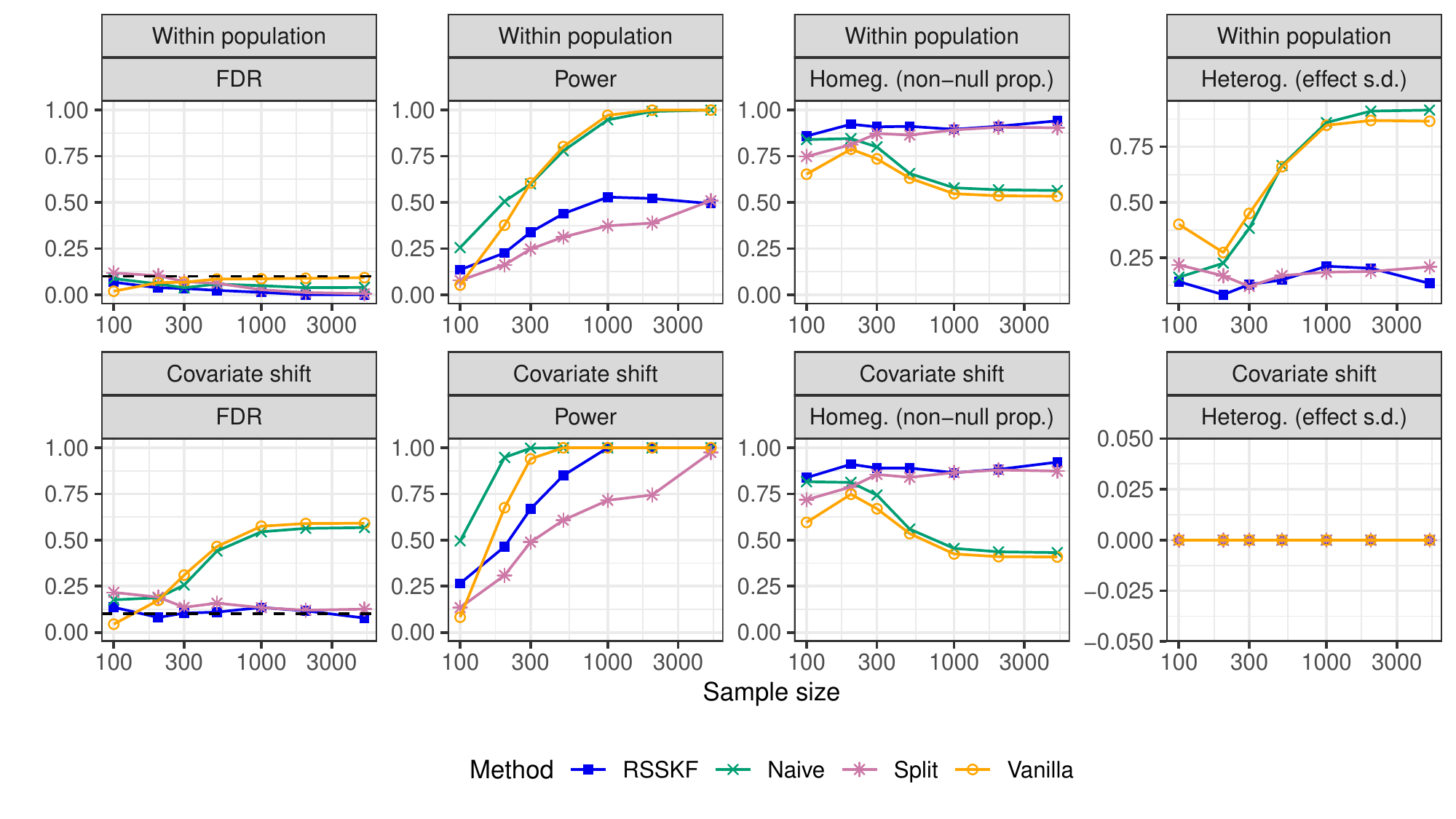}\\[-1em]
  \caption{Performance of the robust subgroup-selective knockoff filter (RSSKF) and alternative benchmarks for subgroup-specific conditional testing under covariate shift, in numerical experiments with synthetic data. The findings summarized in the bottom panel are counted as true if and only if they report a variable whose conditional association with the outcome is robust to changes in the covariate distribution. Other details are as in Figure~\ref{fig:experiment-heterogeneous-1}.}
  \label{fig:experiment-transfer-1}
\end{figure}

Figure~\ref{fig:experiment-transfer-1} reports on the results of 100 independent experiments, as a function of the sample size.
The performance of the robust version of the subgroup-selective knockoff filter is quantified in terms of the empirical false discovery rate, the power, and the homogeneity of the reported findings, separately within the training population and under covariate shift. In the latter case, only the discoveries of variables with a direct conditional association with the outcome (i.e., not mediated by any interactions) are counted as true, while the others are considered false positives.
Therefore, neither the naive benchmark nor the vanilla knockoff filter control this notion of false discovery rate under covariate shift because they tend to report all associated variables, including those that are non-null only thanks to interactions. By contrast, the robust subgroup-selective knockoff filter empirically controls the false discovery rate even under covariate shift.
The data-splitting benchmark here is a modified version of our robust subgroup-selective knockoff filter in which half of the observations are used for partitioning the covariate space, without randomly swapping the knockoffs, and the remaining half are utilized for testing the selected hypotheses.
 The results demonstrate data-splitting is valid under covariate shift but it is not as powerful as our method, especially if the sample size is moderately large.

\FloatBarrier

\subsection{Additional details about the blood donation data} \label{app:data-description}

\begin{table}[H]
  \caption{Randomized treatment-control assignments in a field experiment conducted by~\cite{sun2019mobile} to investigate the effectiveness of different incentives for blood donors.} \vspace{0.5em}
  \label{table:data-blood-groups}
  \centering
  \begin{tabular}{ccccccc}
    \toprule
    \multirow{2}{*}{Assignment} & \multirow{2}{*}{Size} & \multicolumn{5}{c}{Treatments $(X_1,\ldots,X_5)$} \\
    \cmidrule(l{2pt}r{2pt}){3-7}
    & & Reminder & Indiv.~reward & Friend & Group reward & Group gift \\
    \midrule
    Control group     & 14k & 0 & 0 & 0 & 0 & 0 \\
    Treated group 1 & 11k & 1 & 0 & 0 & 0 & 0 \\
    Treated group 2 & 11k & 1 & 1 & 0 & 0 & 0 \\
    Treated group 3 & 11k & 1 & 0 & 1 & 0 & 0 \\
    Treated group 3 & 11k & 1 & 1 & 1 & 0 & 0 \\
    Treated group 4 & 11k & 1 & 1 & 1 & 0 & 0 \\
    Treated group 5 & 11k & 0 & 1 & 1 & 1 & 0 \\
    Treated group 6 & 11k & 1 & 1 & 1 & 0 & 1 \\
    \bottomrule
  \end{tabular}
\end{table}


\begin{table}[H]  \centering
  \caption{Goodness-of-fit-diagnostics for knockoff treatments, for the blood donation data. The means and standard deviations of the knockoffs approximately match those of the corresponding true treatments.} \vspace{0.5em}
  \label{tab:knockoffs-1}
  
\begin{tabular}{rrrrrrrrrrr}
\toprule
\multicolumn{1}{c}{ } & \multicolumn{5}{c}{Treatments} & \multicolumn{5}{c}{Knockoffs} \\
\cmidrule(l{3pt}r{3pt}){2-6} \cmidrule(l{3pt}r{3pt}){7-11}
  & $X_1$ & $X_2$ & $X_3$ & $X_4$ & $X_5$ & $\tilde{X}_1$ & $\tilde{X}_2$ & $\tilde{X}_3$ & $\tilde{X}_4$ & $\tilde{X}_5$\\
\midrule
Mean & 0.825 & 0.412 & 0.550 & 0.138 & 0.138 & 0.822 & 0.411 & 0.549 & 0.139 & 0.140\\
Standard deviation & 0.380 & 0.492 & 0.497 & 0.344 & 0.344 & 0.382 & 0.492 & 0.498 & 0.346 & 0.347\\
\bottomrule
\end{tabular}

\end{table}

\begin{table}[H] \centering
  \caption{Goodness-of-fit-diagnostics for knockoff treatments, for the blood donation data. The pairwise correlations between different treatments approximately match those of the corresponding knockoffs, as well as those between treatments and knockoffs. Smaller values of the diagonal correlation terms in the upper-right block of the correlation matrix (in red) tend to correspond to higher power.} \vspace{0.5em}
  \label{tab:knockoffs-2}
  
\begin{tabular}{lrrrrrrrrrr}
\toprule
\multicolumn{1}{c}{ } & \multicolumn{5}{c}{Treatments} & \multicolumn{5}{c}{Knockoffs} \\
\cmidrule(l{3pt}r{3pt}){2-6} \cmidrule(l{3pt}r{3pt}){7-11}
  & $X_1$ & $X_2$ & $X_3$ & $X_4$ & $X_5$ & $\tilde{X}_1$ & $\tilde{X}_2$ & $\tilde{X}_3$ & $\tilde{X}_4$ & $\tilde{X}_5$\\
\midrule
$X_1$ & 1 & 0.386 & 0.509 & 0.184 & 0.184 & \textcolor{red}{0.701} & 0.385 & 0.508 & 0.185 & 0.186\\
$X_2$ & 0.386 & 1 & 0.197 & -0.335 & 0.477 & 0.389 & \textcolor{red}{0.607} & 0.197 & -0.337 & 0.481\\
$X_3$ & 0.509 & 0.197 & 1 & 0.361 & 0.361 & 0.514 & 0.197 & \textcolor{red}{0.612} & 0.364 & 0.365\\
$X_4$ & 0.184 & -0.335 & 0.361 & 1 & -0.159 & 0.186 & -0.334 & 0.362 & \textcolor{red}{0.622} & -0.161\\
$X_5$ & 0.184 & 0.477 & 0.361 & -0.159 & 1 & 0.186 & 0.478 & 0.362 & -0.161 & \textcolor{red}{0.639}\\
\addlinespace
$\tilde{X}_1$ & \textcolor{white}{0.701} & \textcolor{white}{0.389} & \textcolor{white}{0.514} & \textcolor{white}{0.186} & \textcolor{white}{0.186} & 1 & 0.371 & 0.496 & 0.187 & 0.187\\
$\tilde{X}_2$ & \textcolor{white}{0.385} & \textcolor{white}{0.607} & \textcolor{white}{0.197} & \textcolor{white}{-0.334} & \textcolor{white}{0.478} & 0.371 & 1 & 0.189 & -0.319 & 0.464\\
$\tilde{X}_3$ & \textcolor{white}{0.508} & \textcolor{white}{0.197} & \textcolor{white}{0.612} & \textcolor{white}{0.362} & \textcolor{white}{0.362} & 0.496 & 0.189 & 1 & 0.347 & 0.347\\
$\tilde{X}_4$ & \textcolor{white}{0.185} & \textcolor{white}{-0.337} & \textcolor{white}{0.364} & \textcolor{white}{0.622} & \textcolor{white}{-0.161} & 0.187 & -0.319 & 0.347 & 1 & -0.162\\
$\tilde{X}_5$ & \textcolor{white}{0.186} & \textcolor{white}{0.481} & \textcolor{white}{0.365} & \textcolor{white}{-0.161} & \textcolor{white}{0.639} & 0.187 & 0.464 & 0.347 & -0.162 & 1\\
\bottomrule
\end{tabular}

\end{table}

\FloatBarrier

\subsection{Design of semi-synthetic experiments with blood donation data}  \label{app:analysis-semi-synthetic}

The logistic model used to generate the imaginary donation outcomes considered in Section~\ref{sec:data} is:
\begin{align*}
  \text{logit}\left(\P{Y = 1 \mid X, Z} \right) = \varphi(X,Z),
\end{align*}
with
\begin{align} \label{eq:blood-causal-model}
\begin{split}
\varphi(X,Z)
= & -c + \\
  & - b \cdot \I{\text{male}} + b \cdot \I{\text{married}} + b \cdot \I{\text{resident}} - b \cdot \I{\text{age} < 25} +  \\
  & - b \cdot \I{\text{student}} - b \cdot \I{\text{education} < 16} + b \cdot \I{\text{Rh}^-} - b \cdot \I{\text{blood type} \neq \text{O}} +  \\
  & + a \cdot X_1 \left[ (1-\I{\text{resident}}) + (1-\I{\text{donation within 12 months}}) \right]  + \\
  & + a \cdot X_2 \left[ \I{\text{student}} + \I{\text{student}} \cdot \I{\text{male}} \right]  + \\
  & + a \cdot X_3 \left[ (1-\I{\text{male}}) + (1-\I{\text{student}}) \right]  + \\
  & + a \cdot X_4 \left[ \I{\text{student}}) + \I{\text{male}} \right]  + \\
  & + a \cdot X_5 \left[ (1-\I{\text{education} <16}) \right].
\end{split}
\end{align}
Above, the coefficients are set as $a=0.4$, $b=0.2$, while $c$ is such that half of the individuals in the data set on average receive a simulated $Y=1$.

\clearpage

\subsection{Additional results from experiments with blood donation data}  \label{app:analysis-semi-synthetic-results}

\begin{table}[!htb]
  \caption{Discoveries reported by the subgroup-selective knockoff filter in the analysis of semi-synthetic blood donation data, as in Figure~\ref{fig:experiment-blood}. The sample size is 80,000 and the nominal false discovery rate is 10\%.} \vspace{0.5em}
  \label{table:experiment-blood-disc}
    \centering

\begin{tabular}[t]{ccccccr}
\toprule
Treatment & Subgroup & Truth & W & Homeg. & Heterog. & Samples\\
\midrule
 & Resident : 0 and  Male : 1 & Non-null & 5.94 & 1.00 & 0.19 & 32,550\\

 & Resident : 0 and  Male : 0 & Non-null & 5.47 & 1.00 & 0.19 & 19,866\\

 & Resident : 1 and  Male : 1 & Non-null & 2.55 & 0.63 & 0.19 & 16,037\\

\multirow{-4}{*}{\raggedright\arraybackslash Reminder} & Resident : 1 and  Male : 0 & Non-null & 2.24 & 0.64 & 0.19 & 11,547\\
\cmidrule{1-7}
 & Rh-negative : 1 and  Student : 1 & Non-null & 7.81 & 1.00 & 0.20 & 142\\

\multirow{-2}{*}{\raggedright\arraybackslash Individual reward} & Rh-negative : 0 and  Student : 1 & Non-null & 7.12 & 1.00 & 0.20 & 26,043\\
\cmidrule{1-7}
 & Married : 0 and  Male : 0 & Non-null & 8.54 & 1.00 & 0.00 & 22,547\\

 & Married : 0 and  Male : 1 & Non-null & 4.37 & 1.00 & 0.00 & 34,240\\

\multirow{-3}{*}{\raggedright\arraybackslash Friends request} & Married : 1 and  Male : 0 & Non-null & 3.83 & 1.00 & 0.00 & 8,866\\
\cmidrule{1-7}
Group reward & Student : 1 and  Married : 0 & Non-null & 4.43 & 1.00 & 0.20 & 25,915\\
\bottomrule
\end{tabular}

\end{table}

\FloatBarrier

\section{Proof of Theorem~\ref{thm:coin-flip}} \label{app:proofs}

\begin{proof}[Proof of Theorem~\ref{thm:coin-flip}]
It suffices to prove the flip-sign property, $\mathbf{W} \mid \hat{\psi} \; \oset{d}{=}\; \mathbf{W} \odot \mathbf{U} \mid \hat{\psi}$, as the subsequent corollary about false discovery rate control follows from there immediately as in~\cite{candes2018}.
In fact, the sign-flip property is analogous to the standard one from~\cite{candes2018}, reviewed in Section~\ref{sec:knockoffs-review}, and it holds conditional on $\hat{\psi}$, which fully determines our hypotheses.

We begin by recalling that, by construction, both $\hat{\psi}$ and $\smash{\hat{G}}$ are determined by $[\mathbf{X}, \tilde{\mathbf{X}}]_{\mathrm{swap}(\mathbf{V})},\mathbf{Y},\mathbf{Z}$, as well as (possibly) by some completely independent random noise (e.g., due to a cross-validation procedure).
Therefore, conditional on $[\mathbf{X}, \tilde{\mathbf{X}}]_{\mathrm{swap}(\mathbf{V})},\mathbf{Y},\mathbf{Z}$, we may treat $\hat{\psi}$ and $\smash{\hat{G}}$ as fixed.
Similarly, the test statistics $\mathbf{W}$ are determined by applying a (possibly randomized) function $\boldsymbol{\tau}$ applied to a knockoff-augmented data set $[\mathbf{X}, \tilde{\mathbf{X}}],\mathbf{Y},\mathbf{Z}$.
Since the possible source of randomness in $\boldsymbol{\tau}$ is also independent of everything else, we may ignore it without loss of generality, in order to simplify the notation as much as possible.

With this premise, we note that if follows from \eqref{eq:swap-tau} that, for any fixed $\mathbf{U} \in \{\pm 1\}^{{\hat{G}} \times p}$,
\begin{align*}
  \mathbf{W}([\mathbf{X}, \tilde{\mathbf{X}}]_{\mathrm{swap}(\mathbf{U})},\mathbf{Y},\mathbf{Z})
  =
  \mathbf{U} \odot \mathbf{W}([\mathbf{X}, \tilde{\mathbf{X}}],\mathbf{Y},\mathbf{Z}).
\end{align*}
Further, it follows from Lemma~\ref{eq:lemma-1} that, conditional on $\left( [\mathbf{X}, \tilde{\mathbf{X}}]_{\mathrm{swap}(\mathbf{V})},\mathbf{Y},\mathbf{Z} \right)$,
\begin{align*}
  \mathbf{W}([\mathbf{X}, \tilde{\mathbf{X}}]_{\mathrm{swap}(\mathbf{U})},\mathbf{Y},\mathbf{Z}) 
  \; \oset{d}{=}\;
  \mathbf{W}([\mathbf{X}, \tilde{\mathbf{X}}],\mathbf{Y},\mathbf{Z} ).
\end{align*}
Combining the two equations above implies that, conditional on $\left( [\mathbf{X}, \tilde{\mathbf{X}}]_{\mathrm{swap}(\mathbf{V})},\mathbf{Y},\mathbf{Z} \right)$,
\begin{align*}
  \mathbf{W} \odot \mathbf{U}
  \; \oset{d}{=} \; \mathbf{W}.
\end{align*}
This concludes the proof because $\hat{\psi}$ is a function of $[\mathbf{X}, \tilde{\mathbf{X}}]_{\mathrm{swap}(\mathbf{V})},\mathbf{Y},\mathbf{Z}$.
\end{proof}

\begin{lemma}\label{eq:lemma-1}
For any subset $\mathcal{S} \subseteq \{1,\ldots,{\hat{G}}\}$ of true null hypotheses $\mathcal{H}^{(g)}_{0,j}$ in~\eqref{eq:null-hyp-env},
\begin{align*}
  [\mathbf{X}, \tilde{\mathbf{X}}]_{\mathrm{swap}(\mathcal{S})}
  \,\oset{d}{=}\, [\mathbf{X}, \tilde{\mathbf{X}}]
\end{align*}
conditional on $\left( [\mathbf{X}, \tilde{\mathbf{X}}]_{\mathrm{swap}(\mathbf{V})},\mathbf{Y},\mathbf{Z}\right)$, where $\mathbf{V}$ is the random matrix used to define $\hat{\psi}$.
\end{lemma}

\begin{proof}[Proof of Lemma~\ref{eq:lemma-1}]
We prove the following stronger equality in joint distribution:
\begin{align*}
  \left( [\mathbf{X}, \tilde{\mathbf{X}}]_{\mathrm{swap}(\mathcal{S})}, [\mathbf{X}, \tilde{\mathbf{X}}]_{\mathrm{swap}(\mathbf{V})},\mathbf{Y},\mathbf{Z} \right)
  \,\oset{d}{=}\,
  \left( [\mathbf{X}, \tilde{\mathbf{X}}], [\mathbf{X}, \tilde{\mathbf{X}}]_{\mathrm{swap}(\mathbf{V})},\mathbf{Y},\mathbf{Z} \right).
\end{align*}
By independence, it suffices to establish the above result for a single row of the data matrices:
\begin{align*}
  \left( (X,\tilde{X})_{\mathrm{swap}(s)}, Z, Y , (X,\tilde{X})_{\mathrm{swap}(v)} \right)
  \oset{d}{=} \left( (X,\tilde{X}), Z, Y , (X,\tilde{X})_{\mathrm{swap}(v)} \right).
\end{align*}
By construction of the knockoffs $\tilde{X}$, we know that
\begin{align} \label{eq:lemma-1-eq-1}
(X,\tilde{X})_{\mathrm{swap}(s)} \mid Z \; \oset{d}{=} \; (X,\tilde{X}) \mid Z.
\end{align}
Further, for any fixed swap $s$ and random swap $v$ it holds that $v \odot s \, \oset{d}{=} \, v$, which implies
\begin{align}  \label{eq:lemma-1-eq-2}
  (X,\tilde{X})_{\mathrm{swap}(v)} \mid (X,\tilde{X})_{\mathrm{swap}(s)}, Z \; \oset{d}{=} \; (X,\tilde{X})_{\mathrm{swap}(v)} \mid (X,\tilde{X}), Z.
\end{align}
Combining~\eqref{eq:lemma-1-eq-1} and~\eqref{eq:lemma-1-eq-2} yields:
\begin{align}  \label{eq:lemma-1-eq-3}
  \left( (X,\tilde{X})_{\mathrm{swap}(s)}, (X,\tilde{X})_{\mathrm{swap}(v)}, Z \right) \; \oset{d}{=} \; \left( (X,\tilde{X}),  (X,\tilde{X})_{\mathrm{swap}(v)}, Z \right).
\end{align}
Now, recall that $Y \indep \tilde{X} \mid Z, X$. Further, $Y \indep X_{s} \mid Z, X_{-s}$ for all $X_{s}$ swapped by $s$ because the latter only involves null variables.
Therefore, by the same argument as in the proof of Lemma 3.2 in~\cite{candes2018},
\begin{align}  \label{eq:lemma-1-eq-4}
  Y \mid Z, (X,\tilde{X})_{\mathrm{swap}(s)}, (X,\tilde{X})_{\mathrm{swap}(v)} \; \oset{d}{=} \; Y \mid Z, (X,\tilde{X}), (X,\tilde{X})_{\mathrm{swap}(v)}.
\end{align}
Finally, combining~\eqref{eq:lemma-1-eq-3} with~\eqref{eq:lemma-1-eq-4} gives the desired result.
\end{proof}


\end{document}